\documentclass[aps,prb,twocolumn,showpacs,letterpaper,superscriptaddress]{revtex4-1}

\usepackage{diagbox}
\usepackage{comment}
\usepackage{color}
\usepackage[usenames,dvipsnames,table]{xcolor}
\usepackage{graphics}\usepackage{epsfig}\usepackage{subfigure}
\usepackage{hyperref} 
\usepackage{multirow}
\usepackage{graphicx}
\usepackage{tabularx}
\usepackage{subfigure}

\begin{document}

\title[CDW in single-layer Pb/Ge(111) driven by Pb-substrate exchange interaction]{Charge density wave in single-layer Pb/Ge(111) driven by Pb-substrate exchange interaction.}
\author{Cesare Tresca}
\address{Sorbonne Universit\'e, Institut des Nanosciences de Paris, UMR7588, F-75252, Paris, France}
\address{Dipartimento di Fisica, Sapienza Universit\`a di Roma, 00185 Roma, Italy}
\address{Dipartimento di scienze fisiche e chimiche,  Universit\`a degli studi dell'Aquila, Via Vetoio 10, I-67100 L'Aquila, Italy}
\email[]{cesare.tresca@aquila.infn.it}

\author{Matteo Calandra}
\address{Sorbonne Universit\'e, CNRS, Institut des Nanosciences de Paris, UMR7588, F-75252, Paris, France} 
\address{Dipartimento di Fisica, Universit\`a di Trento,
via Sommarive 14, I-38123 Povo, Italy} 
\email[]{m.calandrabuonaura@unitn.it}
\vspace{10pt}

\begin{abstract}
Single layer Pb on top of (111) surfaces of group IV semiconductors hosts charge density wave and superconductivity depending on the coverage and on the substrate.
These systems are normally considered to be experimental realizations of single band Hubbard models and their properties are mostly investigated using lattice models with frozen structural degrees of freedom, although the reliability of this
approximation is unclear. Here, we consider the case
of Pb/Ge(111) at $1/3$ coverage, for which surface X-ray diffraction and ARPES data are available. By performing 
first principles calculations, we demonstrate that the
non-local exchange between Pb and the substrate drives the system into a $3\times 3$  charge density wave. The electronic structure of this charge ordered phase
is mainly determined by two effects: the magnitude of the Pb distortion and the large spin-orbit coupling.
Finally, we show that the effect applies also to the $3\times 3$ phase of Pb/Si(111) where
the Pb-substrate exchange interaction
increases the bandwidth by more than a factor $1.5$ with respect to DFT+U, in better agreement with STS data.  The delicate interplay between substrate, structural and electronic degrees of freedom invalidates the widespread interpretation available in literature considering these compounds as physical realizations of single band Hubbard models.
\end{abstract}
\maketitle

%
%
%
%

\section{Introduction}
The interplay between charge ordering and electron-electron interactions in two dimensions is a subject of intense research as it is at the heart of the physics of high T$_c$ cuprates superconductors\cite{Ghiringhelli821,Comin390,daSilvaNeto393}, transition metal dichalcogenides\cite{FAZEKAS1980183,PhysRevLett.121.026401,10.1088/2053-1583/ab23c0}, organic conductors\cite{Jerome} and group IV heavy atoms (Pb, Sn) deposed on (111) surfaces of group
IV light semiconductors such as Si or Ge\cite{Carpinelli1996,Carpinelli1997,OTTAVIANO2000L41,PhysRevLett.94.046101}. Its understanding is primordial to unveil quantum phase transitions in nanoscale and strongly correlated systems.

One of the crucial issues in these systems is the range of the electron-electron interaction \cite{MerinoPhysRevLett.99.036404,TerletskaPhysRevB.95.115149}, as several scenarios have been proposed to occur in charge ordered metals depending on the spatial extension of exchange-correlation effects such as $d$ or $p$-wave superconductivity\cite{PhysRevB.63.134421,PhysRevB.73.060503,WolfPhysRevB.98.174515} as well as different kinds of metal-insulator transitions (Mott, Slater)\cite{RevModPhys.70.1039}.
In a Mott or Slater insulator, the gap opening is entirely due to the competition between the electron-electron interaction and the electronic kinetic energy, with practically no influence of the lattice that is considered to be frozen. 
This assumption justifies the use of low-energy Hamiltonian approaches based on the (extended) Hubbard model
with hopping parameters fitted on density functional theory (DFT) calculations
with semilocal kernels. However, it is often unclear to what extent this assumption is valid in real materials
as the disentanglement of electronic and lattice degrees of freedom can be non obvious (as in the case of Mott-Jahn-Teller insulators, for example \cite{Fabrizio_Tosatti, PhysRevLett.121.026401}).

Ideal single-band Hubbard model systems are not so common in nature mainly because real materials have, in most of the case, a multi-band electronic structure with a complicate interplay of localized and delocalized orbitals. Remarkable exceptions are group IV heavy atoms (Pb, Sn) single-layers deposed on (111) surfaces of group IV light semiconductors (Si, Ge).  At room temperature $1/3$ monolayer coverage of Pb or Sn
grown on top of Si(111) or Ge(111) display  $\sqrt{3}\times\sqrt{3}$ reconstructions ($\alpha$ phase). The
three Si (or Ge) dangling bonds in the top substrate layer, next to the
metal ad-atom, are saturated and a free unsaturated electron
is left at each Pb (or Sn) atom, leading to a perfect single band system with a narrow band dispersion. In the case of Pb/Si(111) this single-band has a substantial spin-orbit
splitting as large as 25$\%$ of the band dispersion\cite{PhysRevLett.120.196402}. At low-T both Pb/Si(111) and Pb/Ge(111) undergo a reversible $3\times 3$ reconstruction \cite{PhysRevLett.94.046101,Carpinelli1996}. 

The $\alpha$-Pb/Si(111) phase has been longtime considered a prototype of Mott-Hubbard insulator\cite{PhysRevLett.94.046101,PhysRevB.75.155411}. A large number of papers\cite{PhysRevLett.110.166401,Hansmann2013,PhysRevLett.123.086401,Santoro1998,PhysRevB.59.1891,PhysRevLett.83.1003,Prez2000,Flores2001,PhysRevB.82.035116,PhysRevB.83.041104,10.1038/ncomms2617,PhysRevB.94.224418,PhysRevLett.119.266802} have studied this and similar systems in the approximation where the lattice is frozen,
the substrate is not supposed to play any important role and, in most cases, neglecting spin-orbit. 

Only very recently, the importance of spin-orbit coupling has been
recognized and calculation in the high temperature $\sqrt{3}\times\sqrt{3}$ phase was carried out in Ref.\onlinecite{PhysRevB.94.224418} while both the high and low teperature phases of $\alpha$-Pb/Si(111) have been considered in Ref.\onlinecite{PhysRevLett.120.196402}, where it was shown that the
$\alpha$-Pb/Si(111) is a pseudogapped metal hosting a chiral spin textured Fermi surface.

In Pb/Si(111) at $1/3$ coverage, the DFT+U approximation is able to reproduce the $3\times3$ charge density wave (CDW) formation,  the pseudogap feature in the electronic structure and the STS spectra, but the band dispersion is underestimated of approximately a factor of $2$ as compared to STS data\cite{PhysRevLett.120.196402}. 
Underestimation of the bandwidth is a typical correlation effect often due to an  improper treatment of correlation in semilocal functionals.  Some authors proposed that the intersite V$_{Pb-Pb}$ Coulomb repulsion could be very important 
for gap opening and band dispersion increase\cite{PhysRevB.59.1891,PhysRevB.94.224418,PhysRevLett.123.086401} and the non-locality of the electron-electron interaction within the 
Pb layer has been claimed to be at the origin of non-conventional superconductivity\cite{PhysRevB.63.134421,PhysRevB.73.060503}. However, 
the Pb-Pb distance is approximately 6.7~\AA~and the lateral overlap between different local orbitals on different Pb is weak (the lateral extension of the Wannier functions is about 2.9~\AA\cite{PhysRevB.59.1891,PhysRevB.94.224418,PhysRevB.62.1556}). On the contrary, it is known that the Wannier function is extended to the lower Si substrate atoms up to the second/third Si bilayer\cite{PhysRevB.59.1891,PhysRevB.94.224418,PhysRevB.62.1556}, opening the possibility of  non local Pb/substrate electron-electron interaction, a yet
unexplored effect in all these systems. 
A recent work\cite{PhysRevLett.123.086401} questioned the 
capability of density functional theory based approach to obtain the correct structure for the Pb/Si(111) surface. However, it is difficult to assess the quality of the structural and electronic properties for the $\alpha$-Pb/Si(111) phase both in the high-T and in the low-T $3\times 3$ reconstruction,
due to the extremely small size of the domains that does not allows neither for surface X-ray diffraction (XRD) nor for ARPES measurements. So far, STM provided contradictory claims on the crystal structure\cite{PhysRevLett.94.046101,PhysRevLett.120.196402,PhysRevLett.123.086401}.
\begin{figure}[h]
\centering
\includegraphics[width=\linewidth]{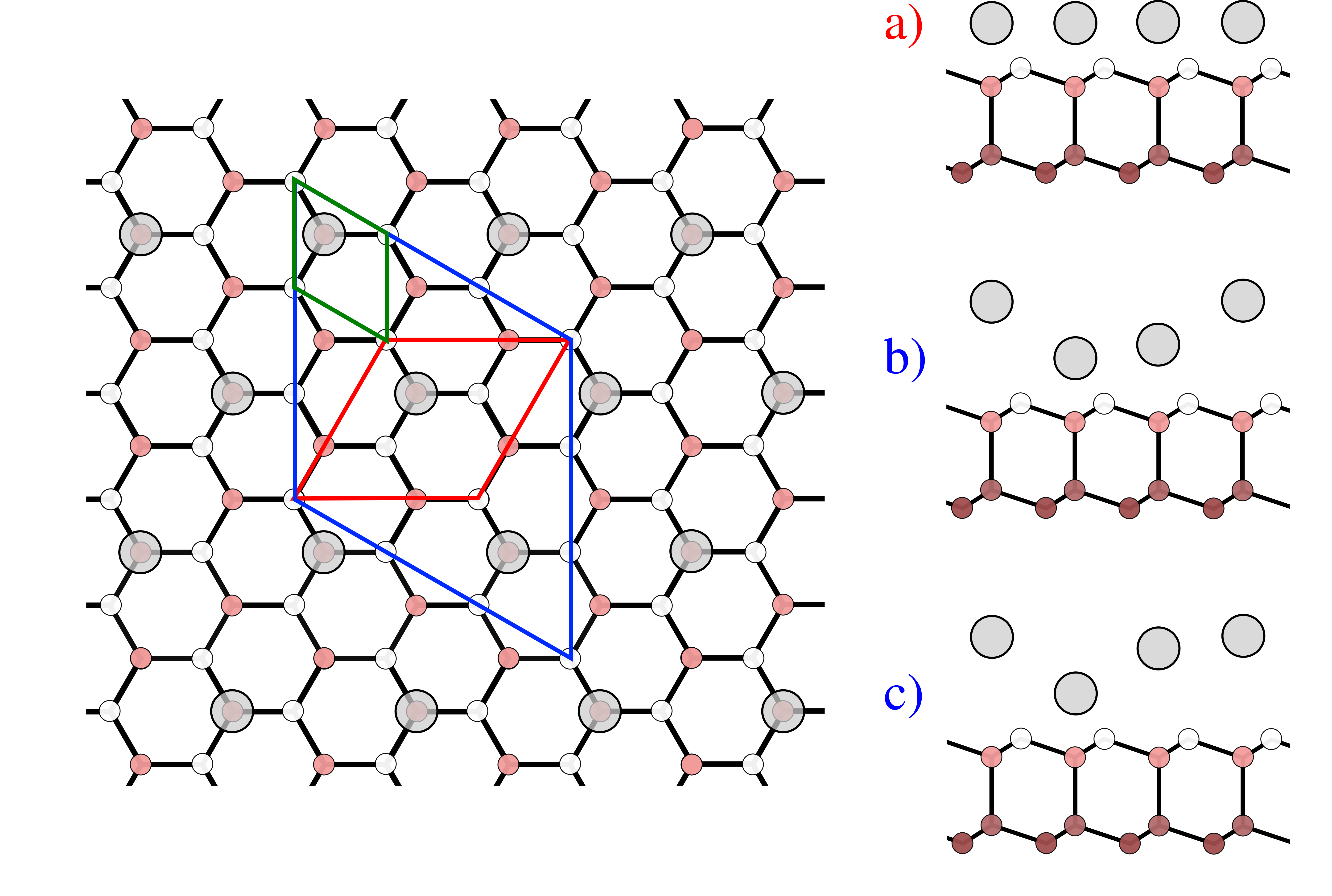}
\caption{Top and lateral view of the $\alpha$-Pb/Ge(111) and
Pb/Si(111) phases. Pb atoms are indicated in light gray, Ge/Si
layers are reported in pink scale from light (topmost layer) to
dark (inner layer). The unit cell of the pristine surface (green),
of the high-T phase (red) and of the low-T one (blue) are
shown. The right panel
is a side view of the three possible
reconstructions: a) $(\sqrt{3}\times\sqrt{3})$-R$30^\circ$; b) $(3\times 3)$ 1up and 2down $(1u2d)$; 
c) $(3\times 3)$ 2up 1down ($1d2u)$.}\label{fig1}
\end{figure}%

The situation is different in Pb/Ge(111) where surface X-ray diffraction \cite{PhysRevLett.82.2524} and ARPES \cite{PhysRevB.57.14758,Tejeda_2007} 
data are available, both for the high-T and low-T phases. Thus, the study of the Pb/Ge(111)
surface at $1/3$ coverage could allow to determine if the structural
properties are correctly reproduced  and if these systems can ultimately be described
in the framework of single band Hubbard or extended Hubbard models with a frozen lattice.
Theory side, few works are present in literature for this system\cite{Carpinelli1996,MASCARAQUE1999337}.

Here we study the Pb/Ge(111) $\sqrt{3}\times\sqrt{3}$ and $3\times3$ reconstructions  as well as the Pb/Si(111) $3\times 3$ reconstruction using range separated hybrid functionals.
By validating first principles calculations against XRD and ARPES, we show that the mechanism for CDW formation in Pb/Ge(111) is not related to an Hubbard like interaction but it is a structural effect driven by the Pb-substrate exchange interaction. Finally we show that the same effect strongly enhances the band dispersion in Pb/Si(111). 

The paper is structured as follows.
After the description of the computational details (Sec. \ref{cdet}), we discuss the results for the high-T $\sqrt{3}\times\sqrt{3}$ phase (Sec. \ref{r3}). Then in Sec. \ref{33} we move to the study of the low-T CDW phase highlighting the key role of Coulomb repulsion and Pb-substrate interaction to properly describe the system. Finally, in Sec. \ref{si} we extend our results on the similar Pb/Si(111) system and in Sec. \ref{conc} we summarize our results and draw our conclusions.

\section {Technical details}\label{cdet}

We model the Pb/Ge(111) and Pb/Si(111) surfaces by considering a layer of Pb atoms on top of 6 Ge or Si bilayers. The bottom dangling bonds are capped with hydrogen.
The atomic position of the first 5 Ge(Si)-substrate layers below the Pb single layer are optimized. The remaining $7$ layers are fixed to the Ge (Si) bulk
  positions. 
The H capping atoms are fixed to the relaxed positions obtained by
capping one 
side of the pristine Ge (Si) surface.
More than 16\AA~of
vacuum are included.
The standard in literature is to use at most three bilayers\cite{PhysRevLett.98.086401,Cudazzo2008747,PhysRevLett.111.106403,PhysRevLett.120.196402}, as this leads to practically converged results for the structural properties of the $\sqrt{3}\times\sqrt{3}$ phase and it is sufficient to return the right physics of the problem in Pb/Si(111). For Pb/Ge(111)  this is not exactly the case.
In fact, we surprisingly find that even if by using 3 bilayers we obtain the right Pb distance from Ge topmost layer in the high-T phase, its high-T electronic properties and the structural properties in the low-T $3\times3$ reconstruction are inaccurate and not converged. In particular, due to an overestimation of the gap related to the modelling with 3 Ge bilayers, this slab is not able to return neither the correct ground state configuration (see Sec.\ref{33}) nor the 
Pb distance from the top of the semiconducting slab, as shown in Tab.\ref{tab0}. This is related to the interaction between
the Pb surface band and the empty conduction states of Ge (Si)  around $\Gamma$. It
is evident from Tab.\ref{tab0} as the 3 Ge bi-layers modelization fails instead the 6
bi-layers one is practically converged.
On the contrary, in the Si case, the Gap is large enough to avoid, or greatly limit, such effects. In fact, for Pb/Si(111) with 6 Si bi-layers we
re-obtain the ”same physics” already described in the literature considering
3 Si bi-layers.
Thus, in the following, we will use a 6 bi-layers substrate that results to be converged with respect to atomic distances and electronic properties around the Fermi level, both in high and low-T phases.

\begin{table}[]
\footnotesize
\begin{center}
Pb/Ge(111):\\
\begin{tabular}{|l|c|c||c|c||c|c|}
\multicolumn{1}{l}{ } &\multicolumn{2}{c}{3 Bi-layers} & \multicolumn{2}{c}{6 Bi-layers}& \multicolumn{2}{c}{9 Bi-layers} \\

\hline
        &    $h$ (\AA) &  $\Delta h$ (\AA)  &     $h$ (\AA)   & $\Delta h$ (\AA)  & $h$ (\AA)   & $\Delta h$ (\AA)  \\ 
\hline
LDA     &        1.88  &    0.01     &        1.89     & 0.18  &  1.89   &  0.18  \\
\hline
GGA     &        1.99  &   $NS$      &        1.97     & 0.21   &  1.99   &  0.20  \\  
\hline
HSE06   &        1.88   &    0.37     &      1.89     & 0.32     &  1.89   & --  \\
\hline
\end{tabular}\\~
\newline
Pb/Si(111):\\
\begin{tabular}{|l|c|c||c|c||c|c|}
\multicolumn{1}{l}{ } &\multicolumn{2}{c}{3 Bi-layers} & \multicolumn{2}{c}{6 Bi-layers} & \multicolumn{2}{c}{9 Bi-layers} \\

\hline
        &    $h$ (\AA) &  $\Delta h$ (\AA)  &     $h$ (\AA)   & $\Delta h$ (\AA) &  $h$ (\AA)   & $\Delta h$ (\AA)  \\ 
\hline
LDA     &           1.87 &      $NS$  &             1.91  &   $NS$  & 1.91    &  $NS$\\
\hline
GGA     &           1.96 &      0.24  &            1.97  &    0.18 &  1.97  &  0.18   \\  
\hline
HSE06   &           1.91 &      0.32  &           1.92   &   0.30   &  --  & -- \\
\hline
\end{tabular}
\end{center}
\caption{
Structural parameters for Pb/Ge(111) and Pb/Si(111)
as a function of the number of layers used for the
substrate modelization: the adsorption 
distance of the Pb ad-atom in the $\sqrt 3 \times \sqrt 3$ phase ($h$) and the
height difference between Pb atoms in the $1u2d$ CDW phase
($\Delta h$).
The acronyms $NS$ means "not stable". 
}
\label{tab0}
\end{table}

Density functional theory calculations are performed with \textsc{Quantum-Espresso}\cite{QEcode, QE-2017} and \textsc{Crystal17}\cite{doi:10.1002/wcms.1360,cryman} codes. 
For plane wave calculations, we used ultrasoft pseudopotentials including 5$d$, 6$s$ and 6$p$ states in valence for lead and 3$d$, 4$s$ and 4$p$ in valence for germanium while for silicon we use the same settings of Ref.\cite{PhysRevLett.120.196402}. We used the Local Density Approximation (LDA), the Generalized Gradient Approximation (GGA) and the GGA+U approximation with
an energy cutoff up to 55 Ry. Integration over Brillouin Zone (BZ) was performed using uniform 10(6)$\times$10(6)$\times$1 Monkhorst and Pack grids\cite{PhysRevB.13.5188} for the $\sqrt{3}\times\sqrt{3}$-R30$^o$(3$\times$3) and a 0.001~Ry Gaussian smearing. 

Hybrid-functionals calculations in plane waves for such a larger number of atoms are hardly feasible. Thus, HSE06\cite{doi:10.1063/1.1564060,doi:10.1063/1.2204597} (non-relativistic) calculations were performed by using the \textsc{Crystal17}~\cite{doi:10.1002/wcms.1360,cryman} code with Gaussian
basis sets. The basis sets used for these calculations have been directly downloaded from the \textsc{Crystal} site. For Pb and Ge we used pseudopotentials\cite{Sophia2013,PhysRevB.74.073101} from the \textsc{Crystal} distribution, while for Si and the capping H an all-electron m-6-311G(d)\cite{doi:10.1063/1.2085170,Pernot2015} and TZVP\cite{Peintinger2012} have been respectively adopted. 
Integration over BZ was performed with the same k-mesh density as in the plane wave calculation and a Fermi-Dirac smearing of 0.0005~Ha. The integration threshold was set to $10^{-7}$ for integrals in the Coulomb series and $10^{-7}$, $10^{-15}$ and $10^{-30}$ for the exchange ones (see Ref.\cite{doi:10.1002/wcms.1360,cryman} for more details).
In this framework, we optimize the internal coordinates. 
We verified that (i) the results with semilocal functionals are consistent with plane waves and Gaussian basis sets and (ii) our \textsc{Crystal17} HSE06 calculations on bulk Ge are compatible with results in literature\cite{PhysRevB.80.115205}. 

Relativistic effects are not implemented in the \textsc{Crystal17} code. To overcome
this difficulty, we fit the non-relativistic 
HSE06 electronic structure and Fermi surfaces at HSE06 fixed geometry in a DFT+U formalism with $U$ on the $l=1$ channel both on Pb and Ge, and then apply non-collinear spin-orbit on top at fixed atomic coordinates  in a DFT+U+SOC calculation. This is possible because relativistic effects are negligible in the atomic relaxation process\cite{PhysRevLett.120.196402}, 
vice versa a good description of the
energy Gap of Ge is necessary for the structural prediction. Large values of $U$ on Pb and substrate states are needed
to reproduce the HSE06 electronic structure on the HSE06 geometry
mainly due to the inability of semilocal and DFT+U functionals in
explaining the gap opening in semiconductive substrate.
Hereafter we will label this approach as relativistic or spin-orbit coupling (SOC).

\section{High-T $\sqrt{3}\times\sqrt{3}$ phase in ${\rm Pb/Ge(111)}$}\label{r3}
\begin{figure}[h]
\centering
\includegraphics[width=0.950\linewidth]{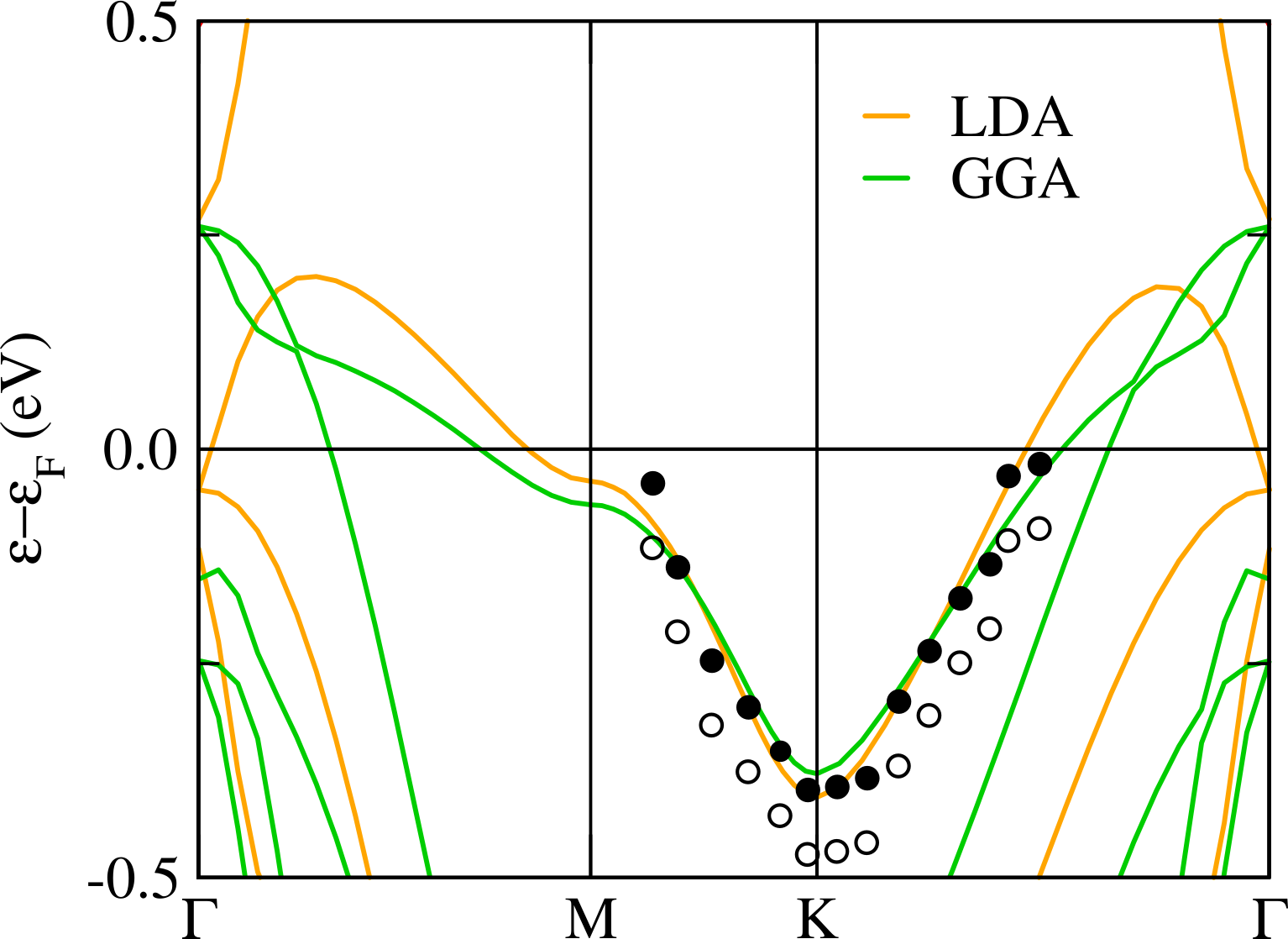}
\caption{Electronic structure of the Pb/Ge(111) $\sqrt{3}\times\sqrt{3}$ phase in the LDA (orange) and GGA (green) approximations.  Black circles are experimental ARPES data from Ref.\cite{PhysRevB.57.14758} (with open circles we report the original data, with filled symbols the shifted ones by +0.075 eV, see main text). 
}\label{fig2}
\end{figure}%

\begin{figure*}[]
\centering
\includegraphics[width=0.45\linewidth]{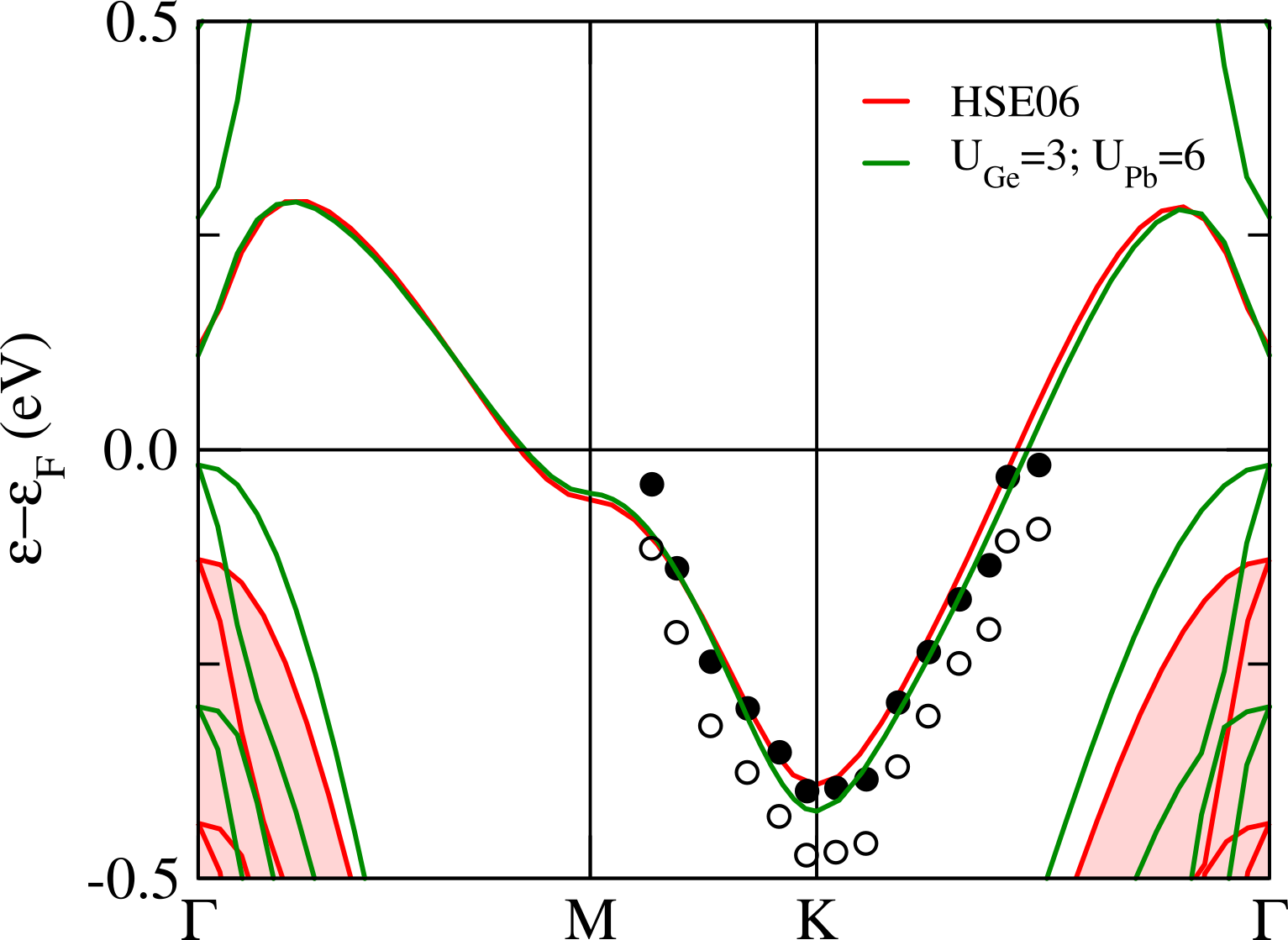}\includegraphics[width=0.45\linewidth]{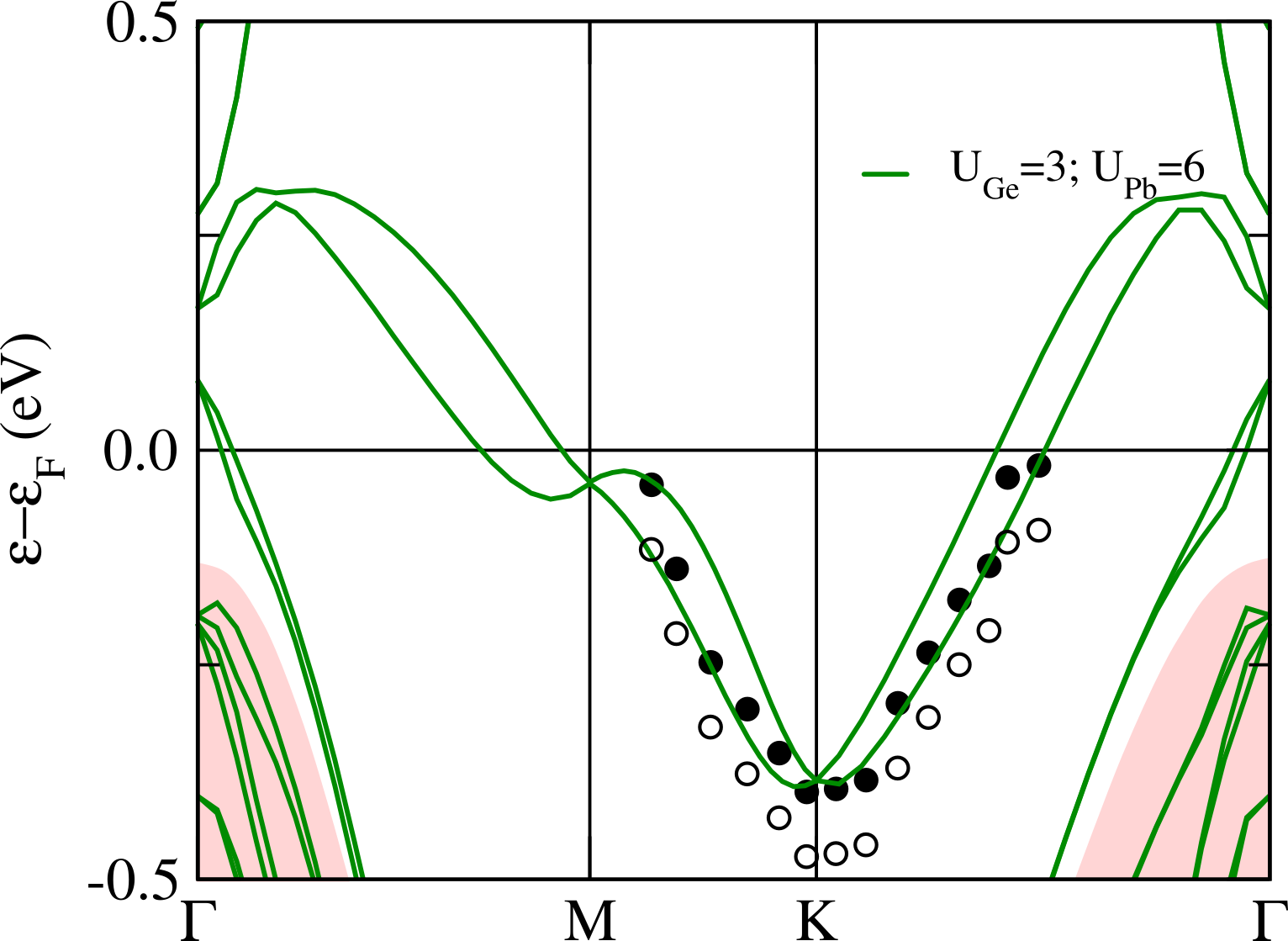}\\
\includegraphics[width=0.49\linewidth]{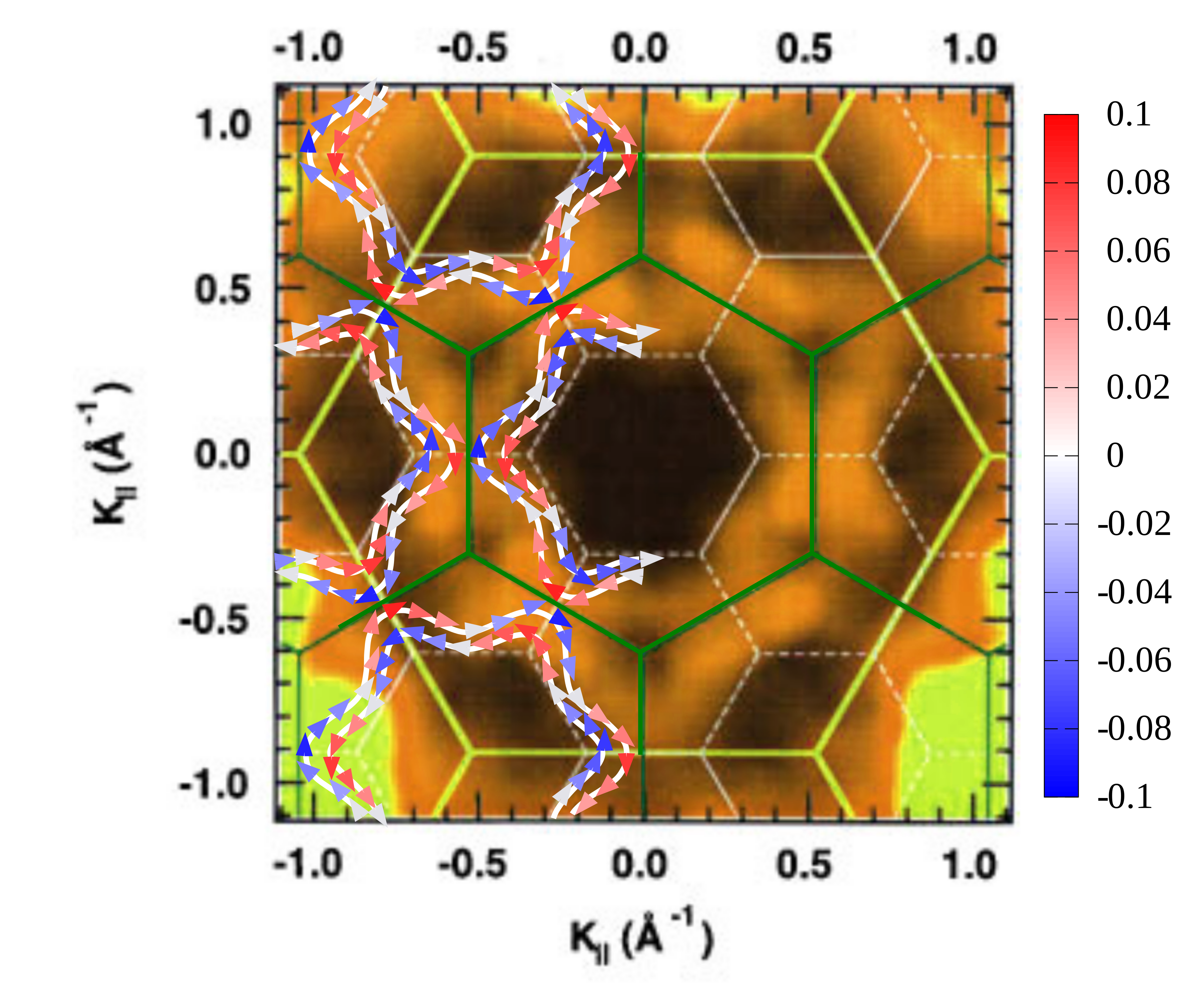}\quad\includegraphics[width=0.49\linewidth]{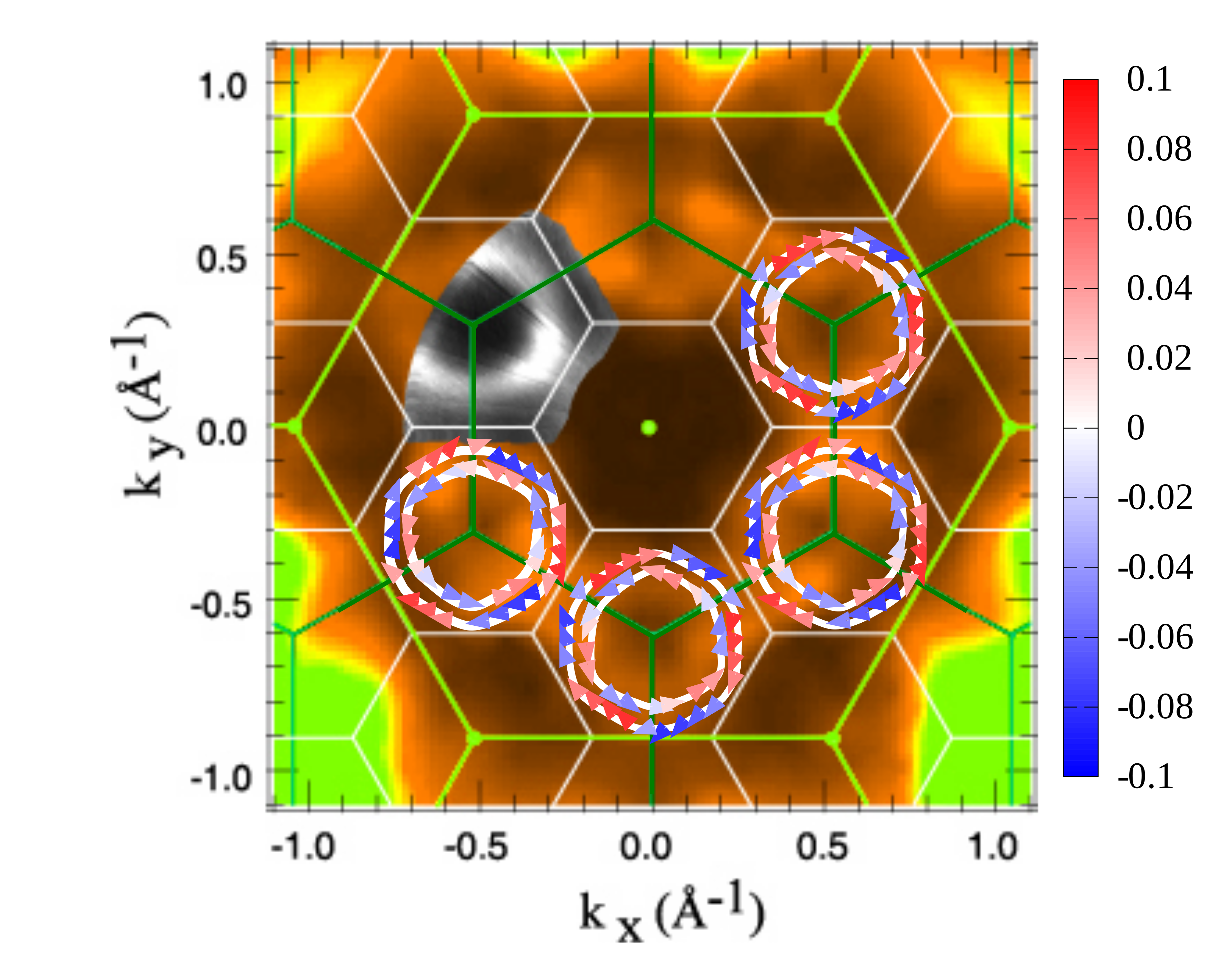}
\caption{Top left (top right) panel: electronic band structure without (with) relativistic effects (see text for more details). Black circles are experimental ARPES data from Ref.\cite{PhysRevB.57.14758} (with open circles we report the original data, with filled symbols the shifted ones, shift of +0.075 eV). The projected Ge bulk HSE06 band structure on the
surface is plotted in light red areas. Bottom panels: calculated Fermi surfaces against ARPES experiments\cite{PhysRevB.57.14758,Tejeda_2007}. In the bottom
right panel we shift the Fermi level in calculations by -0.075 eV to better compare with ARPES\cite{Tejeda_2007} and show the strong sensitivity of the Fermi surface to small Fermi level shifts.  The arrows show the chiral spin direction of the different Fermi sheets. The color label the out-of-plane spin component: white arrows 100\% in-plane polarization; blue and red arrows, opposite out-of-plane components. The out-of-plane spin component is at most 10\% of the in-plane one.}\label{fig3}
\end{figure*}%

We first consider the high-T $\sqrt{3}\times\sqrt{3}$ phase in which all Pb atoms have the same height ($h$) with respect to the last Ge layer, see  Fig. \ref{fig1} (panel a). We find, by using three Ge bilayers, $h=1.88$~\AA~ both in LDA and HSE06, while a larger height of $h=1.99$~\AA~is obtained in GGA. The structure is converged with respect to the slab thickness as in HSE06 by using 6 Ge bilayers we obtain $h=1.89$~\AA, while in order to achieve convergence on the electronic properties, 6 bi
layers are necessary, even in the high-T phase. 

The non-relativistic electronic structure calculated using semilocal LDA and GGA kernels is shown in Fig. \ref{fig2} and displays an entanglement of Ge and Pb bands around the Fermi level, a result due to the underestimation of the Ge gap with these approximations. Even if around the $K$ point
of the BZ theory and experiment seems to be in good agreement
(except for an eventual rigid shift), both LDA and GGA approximations even show an additional Fermi surface around the Brillouin zone center that is incompatible with experimental ARPES data (see Fig. \ref{fig3} and Refs. \cite{PhysRevB.57.14758,Tejeda_2007}). Thus, the LDA and GGA failures in describing the Ge electronic structure  result in a too large hybridization between Ge and the Pb surface state.

The HSE06 approximation captures perfectly the electronic structure measured by ARPES\cite{PhysRevB.57.14758,Tejeda_2007}, as shown
in Fig. \ref{fig3}, provided a small shift of $+0.075$ eV is applied to the experimental ARPES binding energies. We attribute this small
shift  to the difficulty in determining the Fermi level in ARPES at room temperature ($300$K$\approx0.025$eV). A gap is opened in Ge bulk states (see shaded regions) and the hybridization between Ge and Pb states is substantially reduced with respect to the
LDA and GGA case.
The agreement is even better when spin-orbit effects are included in the calculation. 

Finally, the calculated and measured ARPES Fermi surfaces are also shown in Fig. \ref{fig3} for the case of the relativistic theoretical bands with unshifted Fermi level (central panel) or shifted by $-0.075$ eV (bottom panel). The agreement is good, despite the fact that the features in ARPES data are not very sharply defined and the Fermi surface is very sensitive to Fermi level shifts in the $25\div75$ meV range from the Fermi level.

\section{Low-T $3\times 3$ reconstruction in ${\rm Pb/Ge(111)}$}\label{33}

We then move to the low-T charge ordered $3\times 3$ phase, for which surface X-ray diffraction (XRD) experiments\cite{PhysRevLett.82.2524} are available. 
Experiments\cite{Carpinelli1996,PhysRevB.57.14758,MASCARAQUE1999337,Tejeda_2007,PhysRevLett.82.2524} suggest that this low-T phase is characterized by $3\times3$ periodicity with three inequivalent Pb atoms, one of which is higher and the other two lower. For this
reason, this reconstruction is usually labeled $1$ up and $2$ down ($1u2d$) (see Fig. \ref{fig1}, panel b).
As mentioned, the internal coordinates were determined by XRD and
three inequivalent heights ($h$) of the Pb atoms with respect to the topmost Ge layer were detected (see Tab. \ref{tab1}). As it is customary in literature, we label $\Delta h$ the differences between the $z$ coordinates of the topmost Pb with the lower two Pb atoms. In XRD experiments it was found $\Delta h=0.42$ and $0.38$~\AA\cite{PhysRevLett.82.2524} (see Tab. \ref{tab1}).
These measurements are consistent with what was found in the similar Pb/Si(111) system, which in the low-T phase, exhibits a $3\times3$ periodicity with a theoretically predicted $1u2d$ arrangement of the Pb atoms\cite{Cudazzo2008747,PhysRevLett.120.196402}, despite conflicting results in STM data\cite{PhysRevLett.94.046101,PhysRevLett.120.196402,PhysRevLett.123.086401}. 
In fact, a recent work\cite{PhysRevLett.123.086401} claims that the low-T Pb/Si(111) ground state would appear to be the 1 down and 2 up structure ($2u1d$) (see Fig. \ref{fig1}, panel c). For this reason, in the following, we will also study the possible occurrence of the $2u1d$ structure.

\begin{table*}[]
\footnotesize
\begin{center}
\begin{tabular}{|c|c|c|c||c|c|c|}
 \multicolumn{1}{c}{ } &\multicolumn{3}{c}{Energy (eV/Pb)} & \multicolumn{3}{c}{Structural parameters (\AA)}\\\hline
\diagbox[width=9em]{Functional}{System}& $\sqrt{3}\times\sqrt{3}$  & $3\times3$ (1u2d) & $3\times3$ (2u1d) & 
$\Delta h_{\rm Pb-Ge}(\sqrt{3}\times\sqrt{3})$ & $\Delta h_{Pb-Ge}({\rm 1u2d})$& $\Delta h_{\rm Pb}({\rm 1u2d})$\\

\hline
LDA    &  +0.100  &  {\bf 0.000}   & +0.001   &  1.89  &  1.95 (1.97;1.96)  & 0.18 (0.18)    \\ 
\hline
GGA    &  +0.105  & {\bf 0.000}    & {\bf +0.000}  & 1.97  &  2.05 (1.95) & 0.21 (0.21)   \\ 
\hline
HSE06    & +0.046 &  {\bf 0.000}   & +0.093    & 1.89   &  1.99 (1.85)& 0.32 (0.30)  \\ 
\hline
{\bf EXP (XRD)}\cite{PhysRevLett.82.2524} &  $T>250 K$    &   $T<250 K$   & -- & -- & 2.01 (1.92) $\pm 0.05$ & ${\bf 0.42  (0.38)} \pm 0.01$    \\ 
\hline
\end{tabular}
\end{center}
\caption{Stability of the different phases of Pb/Ge(111) using a $6$ Ge bilayers slab. The high-T phase is labeled $\sqrt{3}\times\sqrt{3}$. The experimental stable phase at low-T is  
the 1up 2down $3\times3$ reconstruction ($1u2d$). The energy differences in the first three columns are reported
with respect to the experimental (and theoretical) low-T phase ($1u2d$). In the fourth and fifth column we report the height difference between each Pb atom with respect the Ge layer below in the high-T and low-T phases. 
In the sixth column we show the height difference between the Pb atoms ($\Delta h_{Pb}$) for the $3\times3$ ground state and the comparison with experimental XRD data. As in the $3\times3$ reconstruction the three Pb sites are not equivalent, there are two $\Delta h$ values, the second one is reported in parenthesis. 
}
\label{tab1}
\end{table*}
\begin{figure*}[]
\centering
\includegraphics[height=5.75cm]{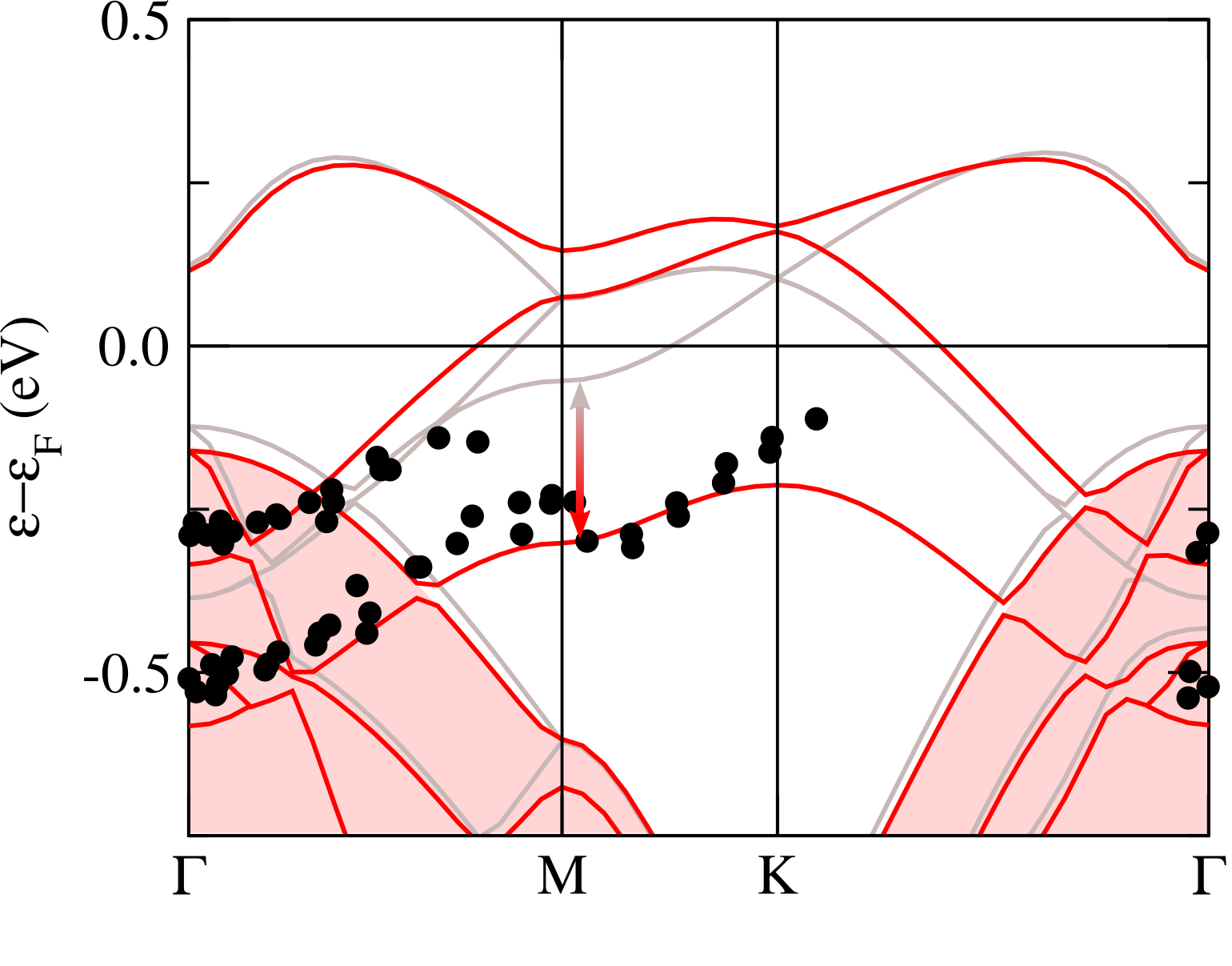}\qquad\includegraphics[height=5.75cm]{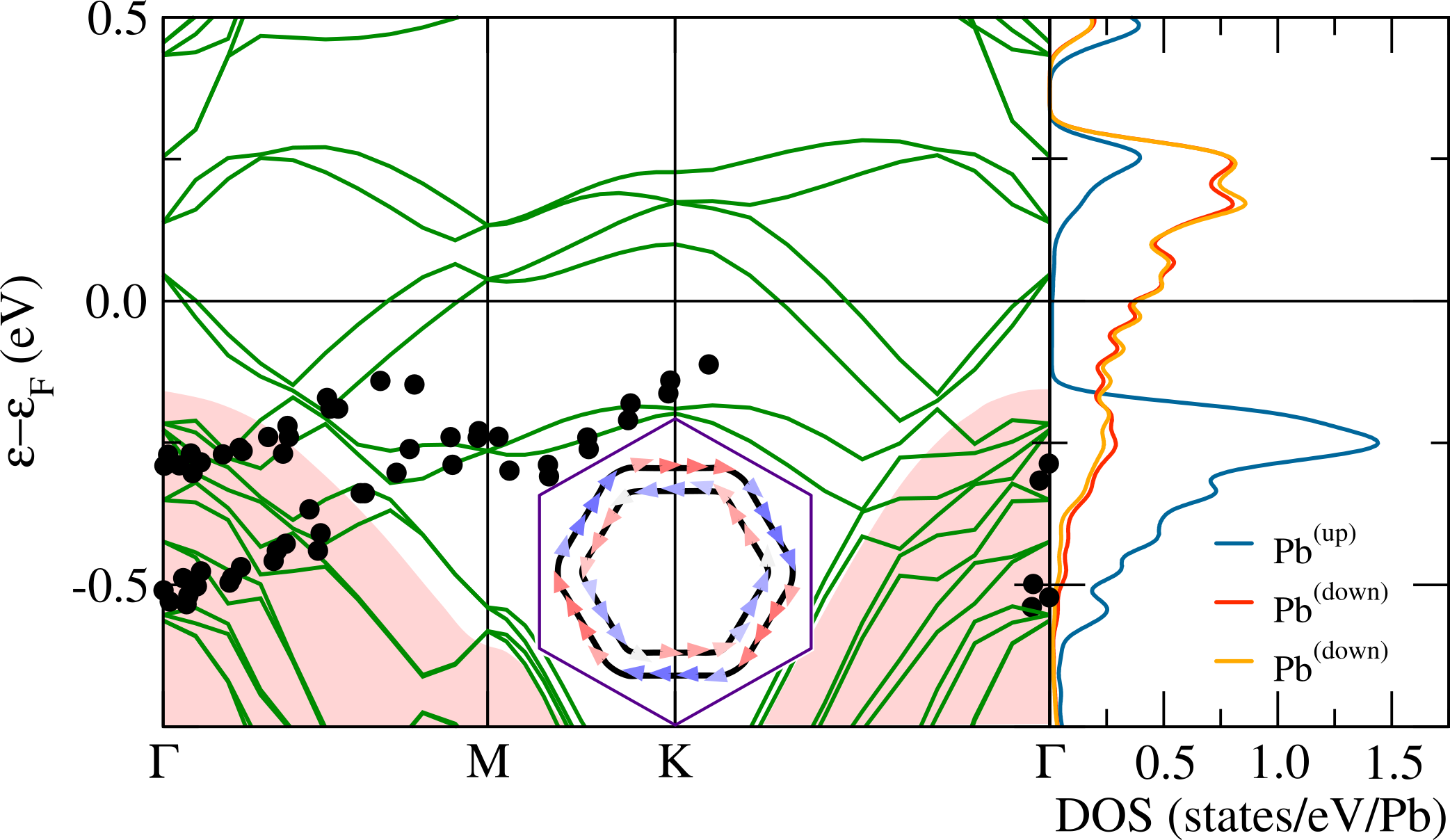}
\caption{In the left panel we report the electronic dispersions obtained for the HSE06 optimized $3\times 3$ (red) and $\sqrt{3}$ (gray) phases plotted in the $3\times 3 $ Brillouin zone. The band deformation labeled by the arrow is proportional to $\Delta h$: it is a structural effect due to the CDW. 
In the right panel the same but with SOC and using the DFT+U approximation
that fits better HSE06 on the $3\times 3$ structure at fixed ions ($U_{Ge}$=3~eV; $U_{Pb}$=6~eV). In black dots we report the ARPES measurements\cite{PhysRevB.57.14758,MASCARAQUE1999337} and in light red the HSE06 bulk bands projected on the surface. 
We report also the density of states projected on Pb atoms highlighting the origin of the different bands. In the inset we plot the Fermi surface of Pb/Ge(111) in the low-T $3\times 3$ phase including relativistic effects. The arrows show the chiral spin direction of the different Fermi sheets. The color label the out-of-plane spin component: white arrows 100\% in-plane polarization; blue and red arrows, opposite out-of-plane components. The out-of-plane spin component is at most 10\% of the in-plane one.
}\label{fig4}
\end{figure*}%
In order to model the transition between the high-T $\sqrt{3}\times\sqrt{3}$ phase and the low-T $3\times 3$ CDW, we perform structural optimization with LDA, GGA and HSE06. 
We find that the energetics, the structural parameters and the stability of the different phases  in the charge ordered $3\times 3$ reconstruction depend on the number of layers used in the calculation (see Tab.\ref{tab0}). We find converged results only
for 6 Ge bilayers, while the popular slab structure with only 3 Ge bilayers leads to incorrect results ($i.e.$ using the 3 Ge bilayers slab in conjunction with the GGA approximation the $2u1d$ structure is erroneously predicted to be the ground state).
The reason is the better description of the substrate energy Gap together with the large Pb-Ge orbital overlap extending deeper in the substrate than for the $\sqrt{3}\times\sqrt{3}$ high-T phase.

The results of the structural optimization are shown in Tab.\ref{tab1}.
Both semilocal LDA, GGA functionals reproduce a $3\times 3$ periodicity for the ground state, as both the $1u2d$ and the $2u1d$ configurations are lower in energy with respect to the $\sqrt{3}\times\sqrt{3}$ phase.
However, in these approximations, the $1u2d$ and the $2u1d$ reconstructions are practically degenerate, in disagreement with experiments finding a $1u2d$ stable in all the sample and excluding the possible coexistence of the two phases\cite{PhysRevLett.82.2524}. 
Finally, the internal parameters, and in particular the  $\Delta h$, turn out to be substantially underestimated (almost a factor of two for the LDA case).

The HSE06 hybrid functional corrects the failure of LDA and GGA and ($i$) succeeds in predicting the stability of the $1u2d$ structure since the $2u1d$ CDW has an even higher energy than the $\sqrt3\times\sqrt3$ structure,
 ($ii$) leads to internal parameters in much better agreement with experimental XRD data.
These results highlight the crucial role played by the non-local exchange interaction between the Pb layer and the Ge substrate in determining the structural properties of the Pb/Ge(111) surface in the charge ordered phase. To our knowledge, this is a poorly explored effect in Pb/Sn single-layers  on top of (111) surfaces of group IV semiconductors, as in practically all existing calculations no geometrical optimization of the substrate is performed with hybrid functionals.

The HSE06 structural data are globally in good agreement with experiments, even if $\Delta h$ is still somewhat underestimated. 
It is interesting to note that even if three different Pb heights are allowed
by symmetry in the simulation, not all functionals lead to three different Pb heights. 
In experiments, the difference between the $z$ component of the coordinates of the
two down Pb atoms is $\approx 0.04$\AA\cite{PhysRevLett.82.2524}.  
This small difference is not present in calculations based on semilocal functionals, 
while it is captured by HSE06, although
slightly underestimated ($\approx 0.02$\AA). This highlights once more the better performance of HSE06 in predicting the structural properties of Pb/Ge(111), mainly because of a better description of the Pb-substrate
exchange interaction.

As hybrid functional tends to stabilize magnetic phases we searched also for possible collinear magnetic solutions, but those are unstable suggesting the system to be non-magnetic.

The electronic structure calculated with the HSE06 functional is shown in Fig. \ref{fig4}. The occupied germanium states are close to the Fermi level at zone center without crossing it.

The
structural deformation which leads to the CDW phase  is accompanied by
a strong charge rearrangement, a reduction in symmetry and the removal
of degeneracy of electronic states, thus three Pb bands result from the lowering of the symmetry in the $3\times3$ CDW phase. One completely filled band (bonding) related to the highest of the three Pb atoms (Pb up), while the other two have components on the two down Pb atoms closer to the Ge surface. Of these two Pb-down bands, one is completely empty (non-bonding) and one crosses the Fermi level (labeled metallic).

The separation between the non-bonding and bonding Pb-bands is directly related to the $\Delta h$ parameter and in the limit of $\Delta h\to 0$ one recovers the degeneracy of the three Pb bands as expected from the calculation for the $\sqrt{3}\times\sqrt{3}$ structure (see the grey line in the left panel of Fig. \ref{fig4}). 
Most important,  the splitting of the bands and the bandwidth in the CDW phase are not due to the Mott-Hubbard interaction, as commonly accepted\cite{PhysRevB.95.195151,PhysRevLett.110.166401,Hansmann2013}, but they are a  structural effect determined
by the the non-locality of the exchange electron-electron interaction between the Pb layer and the Ge substrate.
Finally, we remark that this important result is strictly confirmed by ARPES data as shown in Fig. \ref{fig4}. Indeed  the HSE06 electronic structure is in excellent agreement with ARPES, particularly for the position of the bonding band
that crucially depends on $\Delta h$ revealing a pseudogapped system (see Fig. \ref{fig4} right panel).

The inclusion of SOC produces the expected band splitting reported in Fig.~\ref{fig4} (right panel) and the consequent spin polarized Fermi surface in the inset of Fig.~\ref{fig4}. Thus, as it happens in  Pb/Si(111)\cite{PhysRevLett.120.196402}, the $3\times 3$ phase hosts a chiral Fermi surface composed by two hexagons with different spin polarizations, (the external one rotating clockwise and the internal one counterclockwise). The out-of-plane component is ten times higher in Pb/Ge(111) than in Pb/Si(111).

\section{$3\times 3$ phase of ${\rm Pb/Si(111)}$}\label{si}
The natural question arising is how general this effect is, namely, how much the substrate and the non-locality of the exchange interaction affect the
electronic structure. 

To better clarify this issue,
we consider the $3\times 3$ low-T CDW phase of Pb/Si(111). It is known\cite{Cudazzo2008747,PhysRevLett.120.196402} that the LDA approximation does not stabilize a $3\times 3$ CDW, while the ground state in the  GGA approximation is a $1u2d$ reconstruction with a 
$\Delta h\approx 0.24 $\AA\cite{Cudazzo2008747,PhysRevLett.120.196402}. 
In the DFT+U  approximation, $\Delta h\approx 0.32$\AA\cite{PhysRevLett.120.196402}~and both the pseudogap and the surface band dispersion are roughly underestimated by a factor of $2$ when compared to STS data (see Ref.\cite{PhysRevLett.120.196402} and Fig. \ref{fig5}). A similar result is obtained if HSE06 is used on top of the DFT+U geometry (see supplemental data in Ref.\cite{PhysRevLett.120.196402}). However, as we have seen for the case of Pb/Ge(111), structural optimization with 
HSE06 is crucial to obtain accurate structural parameters and electronic structure for the $3\times 3$ phase.

We then perform structural optimization using the HSE06 functional. We find that the non-local exchange Pb-substrate interaction leads to a substantial structural deformation favouring even more the $3\times 3$ $1u2d$ CDW with an enhanced $\Delta h\approx 0.30$\AA~with respect to semilocal functionals (a value very close to the one obtained in $3\times3$ phase of Pb/Ge(111)). In line with what
happens in the case of Pb/Ge(111), the result of a larger $\Delta h$ is a substantial enhancement of the surface band dispersion (more than a factor 1.5) with respect to the DFT+U, as shown in Fig. \ref{fig5}. The calculated density of states is in much better agreement with
STS data, particularly for the unoccupied states, confirming once more the crucial importance of structural effects on the electronic structure. Even in this case, a fixed atoms approach based on a lattice model is completely inappropriate. Indeed, if the electron-electron interaction is included in the calculation  without allowing the lattice to
relax correspondingly, it is impossible to disentangle the CDW contribution to the gap opening from the one due to the Hubbard-like interactions.

\begin{figure}[]
\centering
\includegraphics[width=0.99\linewidth]{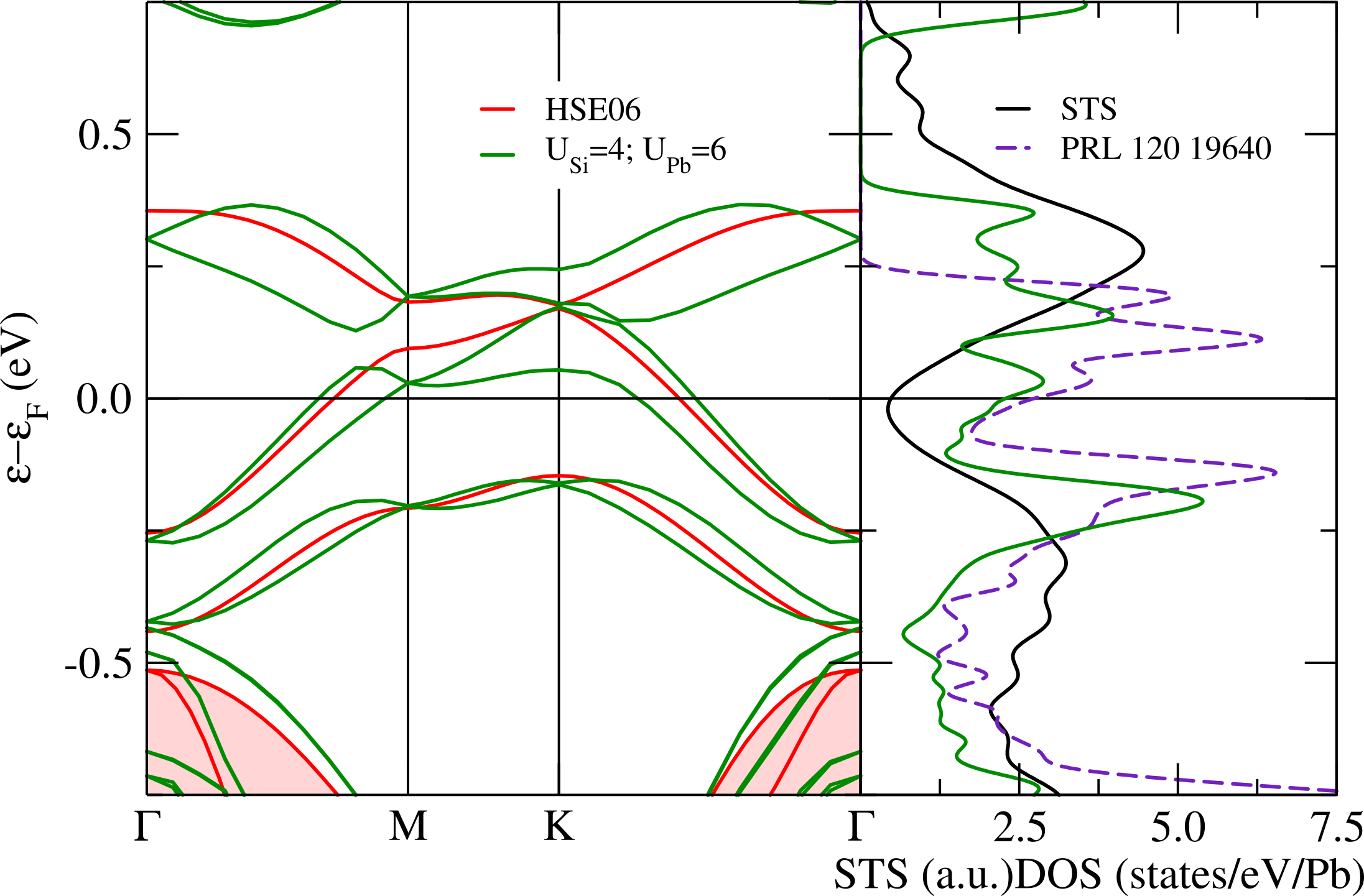}
\caption{Pb/Si(111) electronic band structures obtained with the HSE06 (red) and GGA+U performed on the HSE06 relaxed structure (U$_{\rm Si}=4$; U$_{\rm Pb}=6$~eV) (green) with SOC. In the right panel we report the comparison between the density of states obtained in GGA+U (GGA+U calculation performed on the HSE06 relaxed structure) (green),  experimental STS spectra (black), and the density of states obtained in GGA+U approximation including SOC from Ref.\cite{PhysRevLett.120.196402} (violet dashed).
}\label{fig5}
\end{figure}%

\section{Conclusion}\label{conc}

The range of the electron-electron interaction  has been advocated as a possible mechanism both for charge density wave formation\cite{PhysRevLett.123.086401} and
stabilization of exotic pairing states in single layer Pb \cite{WolfPhysRevB.98.174515} on (111) semiconducting surfaces.
This statement is supported by several studies on this and similar systems
\cite{PhysRevLett.110.166401,Hansmann2013,PhysRevLett.123.086401,Santoro1998,PhysRevB.59.1891,PhysRevLett.83.1003,Prez2000,Flores2001,PhysRevB.82.035116,PhysRevB.83.041104,10.1038/ncomms2617,PhysRevB.94.224418} relying on a frozen atom modeling in the framework of Hubbard and 
extended Hubbard model. The reliability
of these approaches is  unclear, as the lattice degrees
of freedom are frozen, spin-orbit is in most of the case neglected and the adatom-substrate interaction absent.
Moreover, in the well studied system Pb/Si(111) at $1/3$ coverage, different STM/STS experimental papers lead to 
contradictory results for the $3\times 3$ charge density wave structure\cite{PhysRevLett.94.046101,PhysRevLett.120.196402,PhysRevLett.123.086401} and the smallness of the samples does not allow for reliable
XRD and ARPES data.

In an effort to clarify the main mechanism leading to a $3\times 3$
reconstruction in single layer Pb on top of (111) surfaces of group IV semiconductors, we have studied the case of
single layer Pb/Ge(111) at $1/3$ coverage. For this system surface X-ray diffraction \cite{PhysRevLett.82.2524} and ARPES measurements \cite{PhysRevB.57.14758,Tejeda_2007} are available so that a clear benchmark of the accuracy of density-functional based techniques in determining the
structural and electronic is possible. The availability of several experimental data for Pb/Ge(111) contrasts the paucity of theoretical calculations published in literature.

We demonstrated the inadequacy of the GGA and LDA semilocal functionals in describing Pb/Ge(111) both in its high and low-T phases. 
The inclusion of screened exchange corrects this failure and
allows for a sound description of structural and electronic properties in both high-T and low-T phases. 

For the high-T phase, our HSE06 calculation with complete optimization of the crystal structure reproduces all available experiments\cite{PhysRevB.57.14758,Tejeda_2007}. We have shown that, 
as in the case of Pb/Si(111), relativistic effects are large and affect the electronic structure substantially.  The relativistic spin polarized Fermi surface is in excellent agreement with ARPES experiments\cite{PhysRevB.57.14758,Tejeda_2007}.

Our main result is that the non-local exchange between Pb and the Ge(111) substrate drives this system into a $3\times 3$ reconstruction adopting the 1up 2down
geometry, with structural parameters in agreement with experiments\cite{PhysRevLett.82.2524}. Even in this case, semilocal functionals underestimate the magnitude of the structural distortion substantially. 
The electronic structure of this charge ordered phase
is mainly determined by two effects: the magnitude of the Pb distortion and the large spin-orbit coupling. 

Finally, we show that the effect is more general than expected as in the $3\times 3$ phase of Pb/Si(111), 
the Pb-substrate exchange interaction
increases the band dispersion by more than a factor $1.5$ with respect to DFT+U, in better agreement with STS data.  

The delicate interplay between  structural and electronic degrees of freedom in these  compounds invalidates the widespread interpretation available in literature considering them as physical realizations of single band Hubbard models.

\section*{Acknowledgements}

This project has received funding from the European Union’s Horizon 2020 research and innovation program Graphene Flagship under grant agreement No 881603.
Computer facilities were provided by CINES, IDRIS, and CEA
TGCC and PRACE (2017174186). We acknowledge useful discussions with Gianni Profeta, Christophe Brun and Tristan Cren.

\vspace{10mm}

\bibliography{bibliography}{}

\begin{thebibliography}{55}%
\makeatletter
\providecommand \@ifxundefined [1]{%
 \@ifx{#1\undefined}
}%
\providecommand \@ifnum [1]{%
 \ifnum #1\expandafter \@firstoftwo
 \else \expandafter \@secondoftwo
 \fi
}%
\providecommand \@ifx [1]{%
 \ifx #1\expandafter \@firstoftwo
 \else \expandafter \@secondoftwo
 \fi
}%
\providecommand \natexlab [1]{#1}%
\providecommand \enquote  [1]{``#1''}%
\providecommand \bibnamefont  [1]{#1}%
\providecommand \bibfnamefont [1]{#1}%
\providecommand \citenamefont [1]{#1}%
\providecommand \href@noop [0]{\@secondoftwo}%
\providecommand \href [0]{\begingroup \@sanitize@url \@href}%
\providecommand \@href[1]{\@@startlink{#1}\@@href}%
\providecommand \@@href[1]{\endgroup#1\@@endlink}%
\providecommand \@sanitize@url [0]{\catcode `\\12\catcode `\$12\catcode
  `\&12\catcode `\#12\catcode `\^12\catcode `\_12\catcode `\%12\relax}%
\providecommand \@@startlink[1]{}%
\providecommand \@@endlink[0]{}%
\providecommand \url  [0]{\begingroup\@sanitize@url \@url }%
\providecommand \@url [1]{\endgroup\@href {#1}{\urlprefix }}%
\providecommand \urlprefix  [0]{URL }%
\providecommand \Eprint [0]{\href }%
\providecommand \doibase [0]{http://dx.doi.org/}%
\providecommand \selectlanguage [0]{\@gobble}%
\providecommand \bibinfo  [0]{\@secondoftwo}%
\providecommand \bibfield  [0]{\@secondoftwo}%
\providecommand \translation [1]{[#1]}%
\providecommand \BibitemOpen [0]{}%
\providecommand \bibitemStop [0]{}%
\providecommand \bibitemNoStop [0]{.\EOS\space}%
\providecommand \EOS [0]{\spacefactor3000\relax}%
\providecommand \BibitemShut  [1]{\csname bibitem#1\endcsname}%
\let\auto@bib@innerbib\@empty
\bibitem [{\citenamefont {Ghiringhelli}\ \emph {et~al.}(2012)\citenamefont
  {Ghiringhelli}, \citenamefont {Le~Tacon}, \citenamefont {Minola},
  \citenamefont {Blanco-Canosa}, \citenamefont {Mazzoli}, \citenamefont
  {Brookes}, \citenamefont {De~Luca}, \citenamefont {Frano}, \citenamefont
  {Hawthorn}, \citenamefont {He}, \citenamefont {Loew}, \citenamefont {Sala},
  \citenamefont {Peets}, \citenamefont {Salluzzo}, \citenamefont {Schierle},
  \citenamefont {Sutarto}, \citenamefont {Sawatzky}, \citenamefont {Weschke},
  \citenamefont {Keimer},\ and\ \citenamefont {Braicovich}}]{Ghiringhelli821}%
  \BibitemOpen
  \bibfield  {author} {\bibinfo {author} {\bibfnamefont {G.}~\bibnamefont
  {Ghiringhelli}}, \bibinfo {author} {\bibfnamefont {M.}~\bibnamefont
  {Le~Tacon}}, \bibinfo {author} {\bibfnamefont {M.}~\bibnamefont {Minola}},
  \bibinfo {author} {\bibfnamefont {S.}~\bibnamefont {Blanco-Canosa}}, \bibinfo
  {author} {\bibfnamefont {C.}~\bibnamefont {Mazzoli}}, \bibinfo {author}
  {\bibfnamefont {N.~B.}\ \bibnamefont {Brookes}}, \bibinfo {author}
  {\bibfnamefont {G.~M.}\ \bibnamefont {De~Luca}}, \bibinfo {author}
  {\bibfnamefont {A.}~\bibnamefont {Frano}}, \bibinfo {author} {\bibfnamefont
  {D.~G.}\ \bibnamefont {Hawthorn}}, \bibinfo {author} {\bibfnamefont
  {F.}~\bibnamefont {He}}, \bibinfo {author} {\bibfnamefont {T.}~\bibnamefont
  {Loew}}, \bibinfo {author} {\bibfnamefont {M.~M.}\ \bibnamefont {Sala}},
  \bibinfo {author} {\bibfnamefont {D.~C.}\ \bibnamefont {Peets}}, \bibinfo
  {author} {\bibfnamefont {M.}~\bibnamefont {Salluzzo}}, \bibinfo {author}
  {\bibfnamefont {E.}~\bibnamefont {Schierle}}, \bibinfo {author}
  {\bibfnamefont {R.}~\bibnamefont {Sutarto}}, \bibinfo {author} {\bibfnamefont
  {G.~A.}\ \bibnamefont {Sawatzky}}, \bibinfo {author} {\bibfnamefont
  {E.}~\bibnamefont {Weschke}}, \bibinfo {author} {\bibfnamefont
  {B.}~\bibnamefont {Keimer}}, \ and\ \bibinfo {author} {\bibfnamefont
  {L.}~\bibnamefont {Braicovich}},\ }\href {\doibase 10.1126/science.1223532}
  {\bibfield  {journal} {\bibinfo  {journal} {Science}\ }\textbf {\bibinfo
  {volume} {337}},\ \bibinfo {pages} {821} (\bibinfo {year}
  {2012})}\BibitemShut {NoStop}%
\bibitem [{\citenamefont {Comin}\ \emph {et~al.}(2013)\citenamefont {Comin},
  \citenamefont {Frano}, \citenamefont {Yee}, \citenamefont {Yoshida},
  \citenamefont {Eisaki}, \citenamefont {Schierle}, \citenamefont {Weschke},
  \citenamefont {Sutarto}, \citenamefont {He}, \citenamefont {Soumyanarayanan},
  \citenamefont {He}, \citenamefont {Le~Tacon}, \citenamefont {Elfimov},
  \citenamefont {Hoffman}, \citenamefont {Sawatzky}, \citenamefont {Keimer},\
  and\ \citenamefont {Damascelli}}]{Comin390}%
  \BibitemOpen
  \bibfield  {author} {\bibinfo {author} {\bibfnamefont {R.}~\bibnamefont
  {Comin}}, \bibinfo {author} {\bibfnamefont {A.}~\bibnamefont {Frano}},
  \bibinfo {author} {\bibfnamefont {M.~M.}\ \bibnamefont {Yee}}, \bibinfo
  {author} {\bibfnamefont {Y.}~\bibnamefont {Yoshida}}, \bibinfo {author}
  {\bibfnamefont {H.}~\bibnamefont {Eisaki}}, \bibinfo {author} {\bibfnamefont
  {E.}~\bibnamefont {Schierle}}, \bibinfo {author} {\bibfnamefont
  {E.}~\bibnamefont {Weschke}}, \bibinfo {author} {\bibfnamefont
  {R.}~\bibnamefont {Sutarto}}, \bibinfo {author} {\bibfnamefont
  {F.}~\bibnamefont {He}}, \bibinfo {author} {\bibfnamefont {A.}~\bibnamefont
  {Soumyanarayanan}}, \bibinfo {author} {\bibfnamefont {Y.}~\bibnamefont {He}},
  \bibinfo {author} {\bibfnamefont {M.}~\bibnamefont {Le~Tacon}}, \bibinfo
  {author} {\bibfnamefont {I.~S.}\ \bibnamefont {Elfimov}}, \bibinfo {author}
  {\bibfnamefont {J.~E.}\ \bibnamefont {Hoffman}}, \bibinfo {author}
  {\bibfnamefont {G.~A.}\ \bibnamefont {Sawatzky}}, \bibinfo {author}
  {\bibfnamefont {B.}~\bibnamefont {Keimer}}, \ and\ \bibinfo {author}
  {\bibfnamefont {A.}~\bibnamefont {Damascelli}},\ }\href {\doibase
  10.1126/science.1242996} {\bibfield  {journal} {\bibinfo  {journal}
  {Science}\ }\textbf {\bibinfo {volume} {343}},\ \bibinfo {pages} {390}
  (\bibinfo {year} {2013})}\BibitemShut {NoStop}%
\bibitem [{\citenamefont {da~Silva~Neto}\ \emph {et~al.}(2014)\citenamefont
  {da~Silva~Neto}, \citenamefont {Aynajian}, \citenamefont {Frano},
  \citenamefont {Comin}, \citenamefont {Schierle}, \citenamefont {Weschke},
  \citenamefont {Gyenis}, \citenamefont {Wen}, \citenamefont {Schneeloch},
  \citenamefont {Xu}, \citenamefont {Ono}, \citenamefont {Gu}, \citenamefont
  {Le~Tacon},\ and\ \citenamefont {Yazdani}}]{daSilvaNeto393}%
  \BibitemOpen
  \bibfield  {author} {\bibinfo {author} {\bibfnamefont {E.~H.}\ \bibnamefont
  {da~Silva~Neto}}, \bibinfo {author} {\bibfnamefont {P.}~\bibnamefont
  {Aynajian}}, \bibinfo {author} {\bibfnamefont {A.}~\bibnamefont {Frano}},
  \bibinfo {author} {\bibfnamefont {R.}~\bibnamefont {Comin}}, \bibinfo
  {author} {\bibfnamefont {E.}~\bibnamefont {Schierle}}, \bibinfo {author}
  {\bibfnamefont {E.}~\bibnamefont {Weschke}}, \bibinfo {author} {\bibfnamefont
  {A.}~\bibnamefont {Gyenis}}, \bibinfo {author} {\bibfnamefont
  {J.}~\bibnamefont {Wen}}, \bibinfo {author} {\bibfnamefont {J.}~\bibnamefont
  {Schneeloch}}, \bibinfo {author} {\bibfnamefont {Z.}~\bibnamefont {Xu}},
  \bibinfo {author} {\bibfnamefont {S.}~\bibnamefont {Ono}}, \bibinfo {author}
  {\bibfnamefont {G.}~\bibnamefont {Gu}}, \bibinfo {author} {\bibfnamefont
  {M.}~\bibnamefont {Le~Tacon}}, \ and\ \bibinfo {author} {\bibfnamefont
  {A.}~\bibnamefont {Yazdani}},\ }\href {\doibase 10.1126/science.1243479}
  {\bibfield  {journal} {\bibinfo  {journal} {Science}\ }\textbf {\bibinfo
  {volume} {343}},\ \bibinfo {pages} {393} (\bibinfo {year}
  {2014})}\BibitemShut {NoStop}%
\bibitem [{\citenamefont {Fazekas}\ and\ \citenamefont
  {Tosatti}(1980)}]{FAZEKAS1980183}%
  \BibitemOpen
  \bibfield  {author} {\bibinfo {author} {\bibfnamefont {P.}~\bibnamefont
  {Fazekas}}\ and\ \bibinfo {author} {\bibfnamefont {E.}~\bibnamefont
  {Tosatti}},\ }\href {\doibase https://doi.org/10.1016/0378-4363(80)90229-6}
  {\bibfield  {journal} {\bibinfo  {journal} {Physica B+C}\ }\textbf {\bibinfo
  {volume} {99}},\ \bibinfo {pages} {183 } (\bibinfo {year}
  {1980})}\BibitemShut {NoStop}%
\bibitem [{\citenamefont {Calandra}(2018)}]{PhysRevLett.121.026401}%
  \BibitemOpen
  \bibfield  {author} {\bibinfo {author} {\bibfnamefont {M.}~\bibnamefont
  {Calandra}},\ }\href {\doibase 10.1103/PhysRevLett.121.026401} {\bibfield
  {journal} {\bibinfo  {journal} {Phys. Rev. Lett.}\ }\textbf {\bibinfo
  {volume} {121}},\ \bibinfo {pages} {026401} (\bibinfo {year}
  {2018})}\BibitemShut {NoStop}%
\bibitem [{\citenamefont {Tresca}\ and\ \citenamefont
  {Calandra}(2019)}]{10.1088/2053-1583/ab23c0}%
  \BibitemOpen
  \bibfield  {author} {\bibinfo {author} {\bibfnamefont {C.}~\bibnamefont
  {Tresca}}\ and\ \bibinfo {author} {\bibfnamefont {M.}~\bibnamefont
  {Calandra}},\ }\href {\doibase 10.1088/2053-1583/ab23c0} {\bibfield
  {journal} {\bibinfo  {journal} {2D Materials}\ }\textbf {\bibinfo {volume}
  {6}},\ \bibinfo {pages} {035041} (\bibinfo {year} {2019})}\BibitemShut
  {NoStop}%
\bibitem [{\citenamefont {J\'erome}(2004)}]{Jerome}%
  \BibitemOpen
  \bibfield  {author} {\bibinfo {author} {\bibfnamefont {D.}~\bibnamefont
  {J\'erome}},\ }\href {\doibase 10.1021/cr030652g} {\bibfield  {journal}
  {\bibinfo  {journal} {Chemical Reviews}\ }\textbf {\bibinfo {volume} {104}},\
  \bibinfo {pages} {5565} (\bibinfo {year} {2004})}\BibitemShut {NoStop}%
\bibitem [{\citenamefont {Carpinelli}\ \emph {et~al.}(1996)\citenamefont
  {Carpinelli}, \citenamefont {Weitering}, \citenamefont {Plummer},\ and\
  \citenamefont {Stumpf}}]{Carpinelli1996}%
  \BibitemOpen
  \bibfield  {author} {\bibinfo {author} {\bibfnamefont {J.~M.}\ \bibnamefont
  {Carpinelli}}, \bibinfo {author} {\bibfnamefont {H.~H.}\ \bibnamefont
  {Weitering}}, \bibinfo {author} {\bibfnamefont {E.~W.}\ \bibnamefont
  {Plummer}}, \ and\ \bibinfo {author} {\bibfnamefont {R.}~\bibnamefont
  {Stumpf}},\ }\href
  {http://gen.lib.rus.ec/scimag/index.php?s=10.1038/381398a0} {\bibfield
  {journal} {\bibinfo  {journal} {Nature}\ }\textbf {\bibinfo {volume} {381}},\
  \bibinfo {pages} {398} (\bibinfo {year} {1996})}\BibitemShut {NoStop}%
\bibitem [{\citenamefont {Carpinelli}\ \emph {et~al.}(1997)\citenamefont
  {Carpinelli}, \citenamefont {Weitering}, \citenamefont {Bartkowiak},
  \citenamefont {Stumpf},\ and\ \citenamefont {Plummer}}]{Carpinelli1997}%
  \BibitemOpen
  \bibfield  {author} {\bibinfo {author} {\bibfnamefont {J.~M.}\ \bibnamefont
  {Carpinelli}}, \bibinfo {author} {\bibfnamefont {H.~H.}\ \bibnamefont
  {Weitering}}, \bibinfo {author} {\bibfnamefont {M.}~\bibnamefont
  {Bartkowiak}}, \bibinfo {author} {\bibfnamefont {R.}~\bibnamefont {Stumpf}},
  \ and\ \bibinfo {author} {\bibfnamefont {E.~W.}\ \bibnamefont {Plummer}},\
  }\href {\doibase 10.1103/PhysRevLett.79.2859} {\bibfield  {journal} {\bibinfo
   {journal} {Phys. Rev. Lett.}\ }\textbf {\bibinfo {volume} {79}},\ \bibinfo
  {pages} {2859} (\bibinfo {year} {1997})}\BibitemShut {NoStop}%
\bibitem [{\citenamefont {Ottaviano}\ \emph {et~al.}(2000)\citenamefont
  {Ottaviano}, \citenamefont {Crivellari}, \citenamefont {Lozzi},\ and\
  \citenamefont {Santucci}}]{OTTAVIANO2000L41}%
  \BibitemOpen
  \bibfield  {author} {\bibinfo {author} {\bibfnamefont {L.}~\bibnamefont
  {Ottaviano}}, \bibinfo {author} {\bibfnamefont {M.}~\bibnamefont
  {Crivellari}}, \bibinfo {author} {\bibfnamefont {L.}~\bibnamefont {Lozzi}}, \
  and\ \bibinfo {author} {\bibfnamefont {S.}~\bibnamefont {Santucci}},\ }\href
  {\doibase https://doi.org/10.1016/S0039-6028(99)00974-7} {\bibfield
  {journal} {\bibinfo  {journal} {Surface Science}\ }\textbf {\bibinfo {volume}
  {445}},\ \bibinfo {pages} {L41 } (\bibinfo {year} {2000})}\BibitemShut
  {NoStop}%
\bibitem [{\citenamefont {Brihuega}\ \emph {et~al.}(2005)\citenamefont
  {Brihuega}, \citenamefont {Custance}, \citenamefont {P\'erez},\ and\
  \citenamefont {G\'omez-Rodr\'{\i}guez}}]{PhysRevLett.94.046101}%
  \BibitemOpen
  \bibfield  {author} {\bibinfo {author} {\bibfnamefont {I.}~\bibnamefont
  {Brihuega}}, \bibinfo {author} {\bibfnamefont {O.}~\bibnamefont {Custance}},
  \bibinfo {author} {\bibfnamefont {R.}~\bibnamefont {P\'erez}}, \ and\
  \bibinfo {author} {\bibfnamefont {J.~M.}\ \bibnamefont
  {G\'omez-Rodr\'{\i}guez}},\ }\href {\doibase 10.1103/PhysRevLett.94.046101}
  {\bibfield  {journal} {\bibinfo  {journal} {Phys. Rev. Lett.}\ }\textbf
  {\bibinfo {volume} {94}},\ \bibinfo {pages} {046101} (\bibinfo {year}
  {2005})}\BibitemShut {NoStop}%
\bibitem [{\citenamefont {Merino}(2007)}]{MerinoPhysRevLett.99.036404}%
  \BibitemOpen
  \bibfield  {author} {\bibinfo {author} {\bibfnamefont {J.}~\bibnamefont
  {Merino}},\ }\href {\doibase 10.1103/PhysRevLett.99.036404} {\bibfield
  {journal} {\bibinfo  {journal} {Phys. Rev. Lett.}\ }\textbf {\bibinfo
  {volume} {99}},\ \bibinfo {pages} {036404} (\bibinfo {year}
  {2007})}\BibitemShut {NoStop}%
\bibitem [{\citenamefont {Terletska}\ \emph {et~al.}(2017)\citenamefont
  {Terletska}, \citenamefont {Chen},\ and\ \citenamefont
  {Gull}}]{TerletskaPhysRevB.95.115149}%
  \BibitemOpen
  \bibfield  {author} {\bibinfo {author} {\bibfnamefont {H.}~\bibnamefont
  {Terletska}}, \bibinfo {author} {\bibfnamefont {T.}~\bibnamefont {Chen}}, \
  and\ \bibinfo {author} {\bibfnamefont {E.}~\bibnamefont {Gull}},\ }\href
  {\doibase 10.1103/PhysRevB.95.115149} {\bibfield  {journal} {\bibinfo
  {journal} {Phys. Rev. B}\ }\textbf {\bibinfo {volume} {95}},\ \bibinfo
  {pages} {115149} (\bibinfo {year} {2017})}\BibitemShut {NoStop}%
\bibitem [{\citenamefont {Gonz\'alez}\ \emph {et~al.}(2001)\citenamefont
  {Gonz\'alez}, \citenamefont {Guinea},\ and\ \citenamefont
  {Vozmediano}}]{PhysRevB.63.134421}%
  \BibitemOpen
  \bibfield  {author} {\bibinfo {author} {\bibfnamefont {J.}~\bibnamefont
  {Gonz\'alez}}, \bibinfo {author} {\bibfnamefont {F.}~\bibnamefont {Guinea}},
  \ and\ \bibinfo {author} {\bibfnamefont {M.~A.~H.}\ \bibnamefont
  {Vozmediano}},\ }\href {\doibase 10.1103/PhysRevB.63.134421} {\bibfield
  {journal} {\bibinfo  {journal} {Phys. Rev. B}\ }\textbf {\bibinfo {volume}
  {63}},\ \bibinfo {pages} {134421} (\bibinfo {year} {2001})}\BibitemShut
  {NoStop}%
\bibitem [{\citenamefont {Hague}(2006)}]{PhysRevB.73.060503}%
  \BibitemOpen
  \bibfield  {author} {\bibinfo {author} {\bibfnamefont {J.~P.}\ \bibnamefont
  {Hague}},\ }\href {\doibase 10.1103/PhysRevB.73.060503} {\bibfield  {journal}
  {\bibinfo  {journal} {Phys. Rev. B}\ }\textbf {\bibinfo {volume} {73}},\
  \bibinfo {pages} {060503} (\bibinfo {year} {2006})}\BibitemShut {NoStop}%
\bibitem [{\citenamefont {Wolf}\ \emph {et~al.}(2018)\citenamefont {Wolf},
  \citenamefont {Schmidt},\ and\ \citenamefont
  {Rachel}}]{WolfPhysRevB.98.174515}%
  \BibitemOpen
  \bibfield  {author} {\bibinfo {author} {\bibfnamefont {S.}~\bibnamefont
  {Wolf}}, \bibinfo {author} {\bibfnamefont {T.~L.}\ \bibnamefont {Schmidt}}, \
  and\ \bibinfo {author} {\bibfnamefont {S.}~\bibnamefont {Rachel}},\ }\href
  {\doibase 10.1103/PhysRevB.98.174515} {\bibfield  {journal} {\bibinfo
  {journal} {Phys. Rev. B}\ }\textbf {\bibinfo {volume} {98}},\ \bibinfo
  {pages} {174515} (\bibinfo {year} {2018})}\BibitemShut {NoStop}%
\bibitem [{\citenamefont {Imada}\ \emph {et~al.}(1998)\citenamefont {Imada},
  \citenamefont {Fujimori},\ and\ \citenamefont {Tokura}}]{RevModPhys.70.1039}%
  \BibitemOpen
  \bibfield  {author} {\bibinfo {author} {\bibfnamefont {M.}~\bibnamefont
  {Imada}}, \bibinfo {author} {\bibfnamefont {A.}~\bibnamefont {Fujimori}}, \
  and\ \bibinfo {author} {\bibfnamefont {Y.}~\bibnamefont {Tokura}},\ }\href
  {\doibase 10.1103/RevModPhys.70.1039} {\bibfield  {journal} {\bibinfo
  {journal} {Rev. Mod. Phys.}\ }\textbf {\bibinfo {volume} {70}},\ \bibinfo
  {pages} {1039} (\bibinfo {year} {1998})}\BibitemShut {NoStop}%
\bibitem [{\citenamefont {Fabrizio}\ and\ \citenamefont
  {Tosatti}(1997)}]{Fabrizio_Tosatti}%
  \BibitemOpen
  \bibfield  {author} {\bibinfo {author} {\bibfnamefont {M.}~\bibnamefont
  {Fabrizio}}\ and\ \bibinfo {author} {\bibfnamefont {E.}~\bibnamefont
  {Tosatti}},\ }\href {\doibase 10.1103/PhysRevB.55.13465} {\bibfield
  {journal} {\bibinfo  {journal} {Phys. Rev. B}\ }\textbf {\bibinfo {volume}
  {55}},\ \bibinfo {pages} {13465} (\bibinfo {year} {1997})}\BibitemShut
  {NoStop}%
\bibitem [{\citenamefont {Tresca}\ \emph {et~al.}(2018)\citenamefont {Tresca},
  \citenamefont {Brun}, \citenamefont {Bilgeri}, \citenamefont {Menard},
  \citenamefont {Cherkez}, \citenamefont {Federicci}, \citenamefont {Longo},
  \citenamefont {Debontridder}, \citenamefont {D'angelo}, \citenamefont
  {Roditchev}, \citenamefont {Profeta}, \citenamefont {Calandra},\ and\
  \citenamefont {Cren}}]{PhysRevLett.120.196402}%
  \BibitemOpen
  \bibfield  {author} {\bibinfo {author} {\bibfnamefont {C.}~\bibnamefont
  {Tresca}}, \bibinfo {author} {\bibfnamefont {C.}~\bibnamefont {Brun}},
  \bibinfo {author} {\bibfnamefont {T.}~\bibnamefont {Bilgeri}}, \bibinfo
  {author} {\bibfnamefont {G.}~\bibnamefont {Menard}}, \bibinfo {author}
  {\bibfnamefont {V.}~\bibnamefont {Cherkez}}, \bibinfo {author} {\bibfnamefont
  {R.}~\bibnamefont {Federicci}}, \bibinfo {author} {\bibfnamefont
  {D.}~\bibnamefont {Longo}}, \bibinfo {author} {\bibfnamefont
  {F.}~\bibnamefont {Debontridder}}, \bibinfo {author} {\bibfnamefont
  {M.}~\bibnamefont {D'angelo}}, \bibinfo {author} {\bibfnamefont
  {D.}~\bibnamefont {Roditchev}}, \bibinfo {author} {\bibfnamefont
  {G.}~\bibnamefont {Profeta}}, \bibinfo {author} {\bibfnamefont
  {M.}~\bibnamefont {Calandra}}, \ and\ \bibinfo {author} {\bibfnamefont
  {T.}~\bibnamefont {Cren}},\ }\href {\doibase 10.1103/PhysRevLett.120.196402}
  {\bibfield  {journal} {\bibinfo  {journal} {Phys. Rev. Lett.}\ }\textbf
  {\bibinfo {volume} {120}},\ \bibinfo {pages} {196402} (\bibinfo {year}
  {2018})}\BibitemShut {NoStop}%
\bibitem [{\citenamefont {Brihuega}\ \emph {et~al.}(2007)\citenamefont
  {Brihuega}, \citenamefont {Custance}, \citenamefont {Ugeda},\ and\
  \citenamefont {G\'omez-Rodr\'{\i}guez}}]{PhysRevB.75.155411}%
  \BibitemOpen
  \bibfield  {author} {\bibinfo {author} {\bibfnamefont {I.}~\bibnamefont
  {Brihuega}}, \bibinfo {author} {\bibfnamefont {O.}~\bibnamefont {Custance}},
  \bibinfo {author} {\bibfnamefont {M.~M.}\ \bibnamefont {Ugeda}}, \ and\
  \bibinfo {author} {\bibfnamefont {J.~M.}\ \bibnamefont
  {G\'omez-Rodr\'{\i}guez}},\ }\href {\doibase 10.1103/PhysRevB.75.155411}
  {\bibfield  {journal} {\bibinfo  {journal} {Phys. Rev. B}\ }\textbf {\bibinfo
  {volume} {75}},\ \bibinfo {pages} {155411} (\bibinfo {year}
  {2007})}\BibitemShut {NoStop}%
\bibitem [{\citenamefont {Hansmann}\ \emph
  {et~al.}(2013{\natexlab{a}})\citenamefont {Hansmann}, \citenamefont {Ayral},
  \citenamefont {Vaugier}, \citenamefont {Werner},\ and\ \citenamefont
  {Biermann}}]{PhysRevLett.110.166401}%
  \BibitemOpen
  \bibfield  {author} {\bibinfo {author} {\bibfnamefont {P.}~\bibnamefont
  {Hansmann}}, \bibinfo {author} {\bibfnamefont {T.}~\bibnamefont {Ayral}},
  \bibinfo {author} {\bibfnamefont {L.}~\bibnamefont {Vaugier}}, \bibinfo
  {author} {\bibfnamefont {P.}~\bibnamefont {Werner}}, \ and\ \bibinfo {author}
  {\bibfnamefont {S.}~\bibnamefont {Biermann}},\ }\href {\doibase
  10.1103/PhysRevLett.110.166401} {\bibfield  {journal} {\bibinfo  {journal}
  {Phys. Rev. Lett.}\ }\textbf {\bibinfo {volume} {110}},\ \bibinfo {pages}
  {166401} (\bibinfo {year} {2013}{\natexlab{a}})}\BibitemShut {NoStop}%
\bibitem [{\citenamefont {Hansmann}\ \emph
  {et~al.}(2013{\natexlab{b}})\citenamefont {Hansmann}, \citenamefont
  {Vaugier}, \citenamefont {Jiang},\ and\ \citenamefont
  {Biermann}}]{Hansmann2013}%
  \BibitemOpen
  \bibfield  {author} {\bibinfo {author} {\bibfnamefont {P.}~\bibnamefont
  {Hansmann}}, \bibinfo {author} {\bibfnamefont {L.}~\bibnamefont {Vaugier}},
  \bibinfo {author} {\bibfnamefont {H.}~\bibnamefont {Jiang}}, \ and\ \bibinfo
  {author} {\bibfnamefont {S.}~\bibnamefont {Biermann}},\ }\href {\doibase
  10.1088/0953-8984/25/9/094005} {\bibfield  {journal} {\bibinfo  {journal}
  {Journal of Physics: Condensed Matter}\ }\textbf {\bibinfo {volume} {25}},\
  \bibinfo {pages} {094005} (\bibinfo {year} {2013}{\natexlab{b}})}\BibitemShut
  {NoStop}%
\bibitem [{\citenamefont {Adler}\ \emph {et~al.}(2019)\citenamefont {Adler},
  \citenamefont {Rachel}, \citenamefont {Laubach}, \citenamefont {Maklar},
  \citenamefont {Fleszar}, \citenamefont {Sch\"afer},\ and\ \citenamefont
  {Claessen}}]{PhysRevLett.123.086401}%
  \BibitemOpen
  \bibfield  {author} {\bibinfo {author} {\bibfnamefont {F.}~\bibnamefont
  {Adler}}, \bibinfo {author} {\bibfnamefont {S.}~\bibnamefont {Rachel}},
  \bibinfo {author} {\bibfnamefont {M.}~\bibnamefont {Laubach}}, \bibinfo
  {author} {\bibfnamefont {J.}~\bibnamefont {Maklar}}, \bibinfo {author}
  {\bibfnamefont {A.}~\bibnamefont {Fleszar}}, \bibinfo {author} {\bibfnamefont
  {J.}~\bibnamefont {Sch\"afer}}, \ and\ \bibinfo {author} {\bibfnamefont
  {R.}~\bibnamefont {Claessen}},\ }\href {\doibase
  10.1103/PhysRevLett.123.086401} {\bibfield  {journal} {\bibinfo  {journal}
  {Phys. Rev. Lett.}\ }\textbf {\bibinfo {volume} {123}},\ \bibinfo {pages}
  {086401} (\bibinfo {year} {2019})}\BibitemShut {NoStop}%
\bibitem [{\citenamefont {Santoro}\ \emph {et~al.}(1998)\citenamefont
  {Santoro}, \citenamefont {Sorella}, \citenamefont {Becca}, \citenamefont
  {Scandolo},\ and\ \citenamefont {Tossatti}}]{Santoro1998}%
  \BibitemOpen
  \bibfield  {author} {\bibinfo {author} {\bibfnamefont {G.}~\bibnamefont
  {Santoro}}, \bibinfo {author} {\bibfnamefont {S.}~\bibnamefont {Sorella}},
  \bibinfo {author} {\bibfnamefont {F.}~\bibnamefont {Becca}}, \bibinfo
  {author} {\bibfnamefont {S.}~\bibnamefont {Scandolo}}, \ and\ \bibinfo
  {author} {\bibfnamefont {E.}~\bibnamefont {Tossatti}},\ }\href {\doibase
  10.1016/s0039-6028(97)01088-1} {\bibfield  {journal} {\bibinfo  {journal}
  {Surface Science}\ }\textbf {\bibinfo {volume} {402-404}},\ \bibinfo {pages}
  {802} (\bibinfo {year} {1998})}\BibitemShut {NoStop}%
\bibitem [{\citenamefont {Santoro}\ \emph {et~al.}(1999)\citenamefont
  {Santoro}, \citenamefont {Scandolo},\ and\ \citenamefont
  {Tosatti}}]{PhysRevB.59.1891}%
  \BibitemOpen
  \bibfield  {author} {\bibinfo {author} {\bibfnamefont {G.}~\bibnamefont
  {Santoro}}, \bibinfo {author} {\bibfnamefont {S.}~\bibnamefont {Scandolo}}, \
  and\ \bibinfo {author} {\bibfnamefont {E.}~\bibnamefont {Tosatti}},\ }\href
  {\doibase 10.1103/PhysRevB.59.1891} {\bibfield  {journal} {\bibinfo
  {journal} {Phys. Rev. B}\ }\textbf {\bibinfo {volume} {59}},\ \bibinfo
  {pages} {1891} (\bibinfo {year} {1999})}\BibitemShut {NoStop}%
\bibitem [{\citenamefont {Hellberg}\ and\ \citenamefont
  {Erwin}(1999)}]{PhysRevLett.83.1003}%
  \BibitemOpen
  \bibfield  {author} {\bibinfo {author} {\bibfnamefont {C.~S.}\ \bibnamefont
  {Hellberg}}\ and\ \bibinfo {author} {\bibfnamefont {S.~C.}\ \bibnamefont
  {Erwin}},\ }\href {\doibase 10.1103/PhysRevLett.83.1003} {\bibfield
  {journal} {\bibinfo  {journal} {Phys. Rev. Lett.}\ }\textbf {\bibinfo
  {volume} {83}},\ \bibinfo {pages} {1003} (\bibinfo {year}
  {1999})}\BibitemShut {NoStop}%
\bibitem [{\citenamefont {P{\'{e}}rez}\ \emph {et~al.}(2000)\citenamefont
  {P{\'{e}}rez}, \citenamefont {Ortega},\ and\ \citenamefont
  {Flores}}]{Prez2000}%
  \BibitemOpen
  \bibfield  {author} {\bibinfo {author} {\bibfnamefont {R.}~\bibnamefont
  {P{\'{e}}rez}}, \bibinfo {author} {\bibfnamefont {J.}~\bibnamefont {Ortega}},
  \ and\ \bibinfo {author} {\bibfnamefont {F.}~\bibnamefont {Flores}},\ }\href
  {\doibase 10.1016/s0169-4332(00)00418-9} {\bibfield  {journal} {\bibinfo
  {journal} {Applied Surface Science}\ }\textbf {\bibinfo {volume} {166}},\
  \bibinfo {pages} {45} (\bibinfo {year} {2000})}\BibitemShut {NoStop}%
\bibitem [{\citenamefont {Flores}(2001)}]{Flores2001}%
  \BibitemOpen
  \bibfield  {author} {\bibinfo {author} {\bibfnamefont {F.}~\bibnamefont
  {Flores}},\ }\href {\doibase 10.1016/s0079-6816(01)00031-4} {\bibfield
  {journal} {\bibinfo  {journal} {Progress in Surface Science}\ }\textbf
  {\bibinfo {volume} {67}},\ \bibinfo {pages} {299} (\bibinfo {year}
  {2001})}\BibitemShut {NoStop}%
\bibitem [{\citenamefont {Schuwalow}\ \emph {et~al.}(2010)\citenamefont
  {Schuwalow}, \citenamefont {Grieger},\ and\ \citenamefont
  {Lechermann}}]{PhysRevB.82.035116}%
  \BibitemOpen
  \bibfield  {author} {\bibinfo {author} {\bibfnamefont {S.}~\bibnamefont
  {Schuwalow}}, \bibinfo {author} {\bibfnamefont {D.}~\bibnamefont {Grieger}},
  \ and\ \bibinfo {author} {\bibfnamefont {F.}~\bibnamefont {Lechermann}},\
  }\href {\doibase 10.1103/PhysRevB.82.035116} {\bibfield  {journal} {\bibinfo
  {journal} {Phys. Rev. B}\ }\textbf {\bibinfo {volume} {82}},\ \bibinfo
  {pages} {035116} (\bibinfo {year} {2010})}\BibitemShut {NoStop}%
\bibitem [{\citenamefont {Li}\ \emph {et~al.}(2011)\citenamefont {Li},
  \citenamefont {Laubach}, \citenamefont {Fleszar},\ and\ \citenamefont
  {Hanke}}]{PhysRevB.83.041104}%
  \BibitemOpen
  \bibfield  {author} {\bibinfo {author} {\bibfnamefont {G.}~\bibnamefont
  {Li}}, \bibinfo {author} {\bibfnamefont {M.}~\bibnamefont {Laubach}},
  \bibinfo {author} {\bibfnamefont {A.}~\bibnamefont {Fleszar}}, \ and\
  \bibinfo {author} {\bibfnamefont {W.}~\bibnamefont {Hanke}},\ }\href
  {\doibase 10.1103/PhysRevB.83.041104} {\bibfield  {journal} {\bibinfo
  {journal} {Phys. Rev. B}\ }\textbf {\bibinfo {volume} {83}},\ \bibinfo
  {pages} {041104} (\bibinfo {year} {2011})}\BibitemShut {NoStop}%
\bibitem [{\citenamefont {Li}\ \emph {et~al.}(2013)\citenamefont {Li},
  \citenamefont {H\"opfner}, \citenamefont {Sch\"afer}, \citenamefont
  {Blumenstein}, \citenamefont {Meyer}, \citenamefont {Bostwick}, \citenamefont
  {Rotenberg}, \citenamefont {Claessen},\ and\ \citenamefont
  {Hanke}}]{10.1038/ncomms2617}%
  \BibitemOpen
  \bibfield  {author} {\bibinfo {author} {\bibfnamefont {G.}~\bibnamefont
  {Li}}, \bibinfo {author} {\bibfnamefont {P.}~\bibnamefont {H\"opfner}},
  \bibinfo {author} {\bibfnamefont {J.}~\bibnamefont {Sch\"afer}}, \bibinfo
  {author} {\bibfnamefont {C.}~\bibnamefont {Blumenstein}}, \bibinfo {author}
  {\bibfnamefont {S.}~\bibnamefont {Meyer}}, \bibinfo {author} {\bibfnamefont
  {A.}~\bibnamefont {Bostwick}}, \bibinfo {author} {\bibfnamefont
  {E.}~\bibnamefont {Rotenberg}}, \bibinfo {author} {\bibfnamefont
  {R.}~\bibnamefont {Claessen}}, \ and\ \bibinfo {author} {\bibfnamefont
  {W.}~\bibnamefont {Hanke}},\ }\href
  {http://gen.lib.rus.ec/scimag/index.php?s=10.1038/ncomms2617} {\bibfield
  {journal} {\bibinfo  {journal} {Nat Commun}\ }\textbf {\bibinfo {volume}
  {4}},\ \bibinfo {pages} {1620} (\bibinfo {year} {2013})}\BibitemShut
  {NoStop}%
\bibitem [{\citenamefont {Badrtdinov}\ \emph {et~al.}(2016)\citenamefont
  {Badrtdinov}, \citenamefont {Nikolaev}, \citenamefont {Katsnelson},\ and\
  \citenamefont {Mazurenko}}]{PhysRevB.94.224418}%
  \BibitemOpen
  \bibfield  {author} {\bibinfo {author} {\bibfnamefont {D.~I.}\ \bibnamefont
  {Badrtdinov}}, \bibinfo {author} {\bibfnamefont {S.~A.}\ \bibnamefont
  {Nikolaev}}, \bibinfo {author} {\bibfnamefont {M.~I.}\ \bibnamefont
  {Katsnelson}}, \ and\ \bibinfo {author} {\bibfnamefont {V.~V.}\ \bibnamefont
  {Mazurenko}},\ }\href {\doibase 10.1103/PhysRevB.94.224418} {\bibfield
  {journal} {\bibinfo  {journal} {Phys. Rev. B}\ }\textbf {\bibinfo {volume}
  {94}},\ \bibinfo {pages} {224418} (\bibinfo {year} {2016})}\BibitemShut
  {NoStop}%
\bibitem [{\citenamefont {Ming}\ \emph {et~al.}(2017)\citenamefont {Ming},
  \citenamefont {Johnston}, \citenamefont {Mulugeta}, \citenamefont {Smith},
  \citenamefont {Vilmercati}, \citenamefont {Lee}, \citenamefont {Maier},
  \citenamefont {Snijders},\ and\ \citenamefont
  {Weitering}}]{PhysRevLett.119.266802}%
  \BibitemOpen
  \bibfield  {author} {\bibinfo {author} {\bibfnamefont {F.}~\bibnamefont
  {Ming}}, \bibinfo {author} {\bibfnamefont {S.}~\bibnamefont {Johnston}},
  \bibinfo {author} {\bibfnamefont {D.}~\bibnamefont {Mulugeta}}, \bibinfo
  {author} {\bibfnamefont {T.~S.}\ \bibnamefont {Smith}}, \bibinfo {author}
  {\bibfnamefont {P.}~\bibnamefont {Vilmercati}}, \bibinfo {author}
  {\bibfnamefont {G.}~\bibnamefont {Lee}}, \bibinfo {author} {\bibfnamefont
  {T.~A.}\ \bibnamefont {Maier}}, \bibinfo {author} {\bibfnamefont {P.~C.}\
  \bibnamefont {Snijders}}, \ and\ \bibinfo {author} {\bibfnamefont {H.~H.}\
  \bibnamefont {Weitering}},\ }\href {\doibase 10.1103/PhysRevLett.119.266802}
  {\bibfield  {journal} {\bibinfo  {journal} {Phys. Rev. Lett.}\ }\textbf
  {\bibinfo {volume} {119}},\ \bibinfo {pages} {266802} (\bibinfo {year}
  {2017})}\BibitemShut {NoStop}%
\bibitem [{\citenamefont {Profeta}\ \emph {et~al.}(2000)\citenamefont
  {Profeta}, \citenamefont {Continenza}, \citenamefont {Ottaviano},
  \citenamefont {Mannstadt},\ and\ \citenamefont {Freeman}}]{PhysRevB.62.1556}%
  \BibitemOpen
  \bibfield  {author} {\bibinfo {author} {\bibfnamefont {G.}~\bibnamefont
  {Profeta}}, \bibinfo {author} {\bibfnamefont {A.}~\bibnamefont {Continenza}},
  \bibinfo {author} {\bibfnamefont {L.}~\bibnamefont {Ottaviano}}, \bibinfo
  {author} {\bibfnamefont {W.}~\bibnamefont {Mannstadt}}, \ and\ \bibinfo
  {author} {\bibfnamefont {A.~J.}\ \bibnamefont {Freeman}},\ }\href {\doibase
  10.1103/PhysRevB.62.1556} {\bibfield  {journal} {\bibinfo  {journal} {Phys.
  Rev. B}\ }\textbf {\bibinfo {volume} {62}},\ \bibinfo {pages} {1556}
  (\bibinfo {year} {2000})}\BibitemShut {NoStop}%
\bibitem [{\citenamefont {Mascaraque}\ \emph
  {et~al.}(1999{\natexlab{a}})\citenamefont {Mascaraque}, \citenamefont
  {Avila}, \citenamefont {Alvarez}, \citenamefont {Asensio}, \citenamefont
  {Ferrer},\ and\ \citenamefont {Michel}}]{PhysRevLett.82.2524}%
  \BibitemOpen
  \bibfield  {author} {\bibinfo {author} {\bibfnamefont {A.}~\bibnamefont
  {Mascaraque}}, \bibinfo {author} {\bibfnamefont {J.}~\bibnamefont {Avila}},
  \bibinfo {author} {\bibfnamefont {J.}~\bibnamefont {Alvarez}}, \bibinfo
  {author} {\bibfnamefont {M.~C.}\ \bibnamefont {Asensio}}, \bibinfo {author}
  {\bibfnamefont {S.}~\bibnamefont {Ferrer}}, \ and\ \bibinfo {author}
  {\bibfnamefont {E.~G.}\ \bibnamefont {Michel}},\ }\href {\doibase
  10.1103/PhysRevLett.82.2524} {\bibfield  {journal} {\bibinfo  {journal}
  {Phys. Rev. Lett.}\ }\textbf {\bibinfo {volume} {82}},\ \bibinfo {pages}
  {2524} (\bibinfo {year} {1999}{\natexlab{a}})}\BibitemShut {NoStop}%
\bibitem [{\citenamefont {Mascaraque}\ \emph {et~al.}(1998)\citenamefont
  {Mascaraque}, \citenamefont {Avila}, \citenamefont {Michel},\ and\
  \citenamefont {Asensio}}]{PhysRevB.57.14758}%
  \BibitemOpen
  \bibfield  {author} {\bibinfo {author} {\bibfnamefont {A.}~\bibnamefont
  {Mascaraque}}, \bibinfo {author} {\bibfnamefont {J.}~\bibnamefont {Avila}},
  \bibinfo {author} {\bibfnamefont {E.~G.}\ \bibnamefont {Michel}}, \ and\
  \bibinfo {author} {\bibfnamefont {M.~C.}\ \bibnamefont {Asensio}},\ }\href
  {\doibase 10.1103/PhysRevB.57.14758} {\bibfield  {journal} {\bibinfo
  {journal} {Phys. Rev. B}\ }\textbf {\bibinfo {volume} {57}},\ \bibinfo
  {pages} {14758} (\bibinfo {year} {1998})}\BibitemShut {NoStop}%
\bibitem [{\citenamefont {Tejeda}\ \emph {et~al.}(2007)\citenamefont {Tejeda},
  \citenamefont {Cort{\'{e}}s}, \citenamefont {Lobo}, \citenamefont {Michel},\
  and\ \citenamefont {Mascaraque}}]{Tejeda_2007}%
  \BibitemOpen
  \bibfield  {author} {\bibinfo {author} {\bibfnamefont {A.}~\bibnamefont
  {Tejeda}}, \bibinfo {author} {\bibfnamefont {R.}~\bibnamefont
  {Cort{\'{e}}s}}, \bibinfo {author} {\bibfnamefont {J.}~\bibnamefont {Lobo}},
  \bibinfo {author} {\bibfnamefont {E.~G.}\ \bibnamefont {Michel}}, \ and\
  \bibinfo {author} {\bibfnamefont {A.}~\bibnamefont {Mascaraque}},\ }\href
  {\doibase 10.1088/0953-8984/19/35/355008} {\bibfield  {journal} {\bibinfo
  {journal} {Journal of Physics: Condensed Matter}\ }\textbf {\bibinfo {volume}
  {19}},\ \bibinfo {pages} {355008} (\bibinfo {year} {2007})}\BibitemShut
  {NoStop}%
\bibitem [{\citenamefont {Mascaraque}\ \emph
  {et~al.}(1999{\natexlab{b}})\citenamefont {Mascaraque}, \citenamefont
  {Avila}, \citenamefont {Asensio},\ and\ \citenamefont
  {Michel}}]{MASCARAQUE1999337}%
  \BibitemOpen
  \bibfield  {author} {\bibinfo {author} {\bibfnamefont {A.}~\bibnamefont
  {Mascaraque}}, \bibinfo {author} {\bibfnamefont {J.}~\bibnamefont {Avila}},
  \bibinfo {author} {\bibfnamefont {M.}~\bibnamefont {Asensio}}, \ and\
  \bibinfo {author} {\bibfnamefont {E.}~\bibnamefont {Michel}},\ }\href
  {\doibase https://doi.org/10.1016/S0039-6028(99)00129-6} {\bibfield
  {journal} {\bibinfo  {journal} {Surface Science}\ }\textbf {\bibinfo {volume}
  {433-435}},\ \bibinfo {pages} {337 } (\bibinfo {year}
  {1999}{\natexlab{b}})}\BibitemShut {NoStop}%
\bibitem [{\citenamefont {Profeta}\ and\ \citenamefont
  {Tosatti}(2007)}]{PhysRevLett.98.086401}%
  \BibitemOpen
  \bibfield  {author} {\bibinfo {author} {\bibfnamefont {G.}~\bibnamefont
  {Profeta}}\ and\ \bibinfo {author} {\bibfnamefont {E.}~\bibnamefont
  {Tosatti}},\ }\href {\doibase 10.1103/PhysRevLett.98.086401} {\bibfield
  {journal} {\bibinfo  {journal} {Phys. Rev. Lett.}\ }\textbf {\bibinfo
  {volume} {98}},\ \bibinfo {pages} {086401} (\bibinfo {year}
  {2007})}\BibitemShut {NoStop}%
\bibitem [{\citenamefont {Cudazzo}\ \emph {et~al.}(2008)\citenamefont
  {Cudazzo}, \citenamefont {Profeta},\ and\ \citenamefont
  {Continenza}}]{Cudazzo2008747}%
  \BibitemOpen
  \bibfield  {author} {\bibinfo {author} {\bibfnamefont {P.}~\bibnamefont
  {Cudazzo}}, \bibinfo {author} {\bibfnamefont {G.}~\bibnamefont {Profeta}}, \
  and\ \bibinfo {author} {\bibfnamefont {A.}~\bibnamefont {Continenza}},\
  }\href {\doibase http://dx.doi.org/10.1016/j.susc.2007.12.001} {\bibfield
  {journal} {\bibinfo  {journal} {Surface Science}\ }\textbf {\bibinfo {volume}
  {602}},\ \bibinfo {pages} {747 } (\bibinfo {year} {2008})}\BibitemShut
  {NoStop}%
\bibitem [{\citenamefont {Lee}\ \emph {et~al.}(2013)\citenamefont {Lee},
  \citenamefont {Kim},\ and\ \citenamefont {Cho}}]{PhysRevLett.111.106403}%
  \BibitemOpen
  \bibfield  {author} {\bibinfo {author} {\bibfnamefont {J.-H.}\ \bibnamefont
  {Lee}}, \bibinfo {author} {\bibfnamefont {H.-J.}\ \bibnamefont {Kim}}, \ and\
  \bibinfo {author} {\bibfnamefont {J.-H.}\ \bibnamefont {Cho}},\ }\href
  {\doibase 10.1103/PhysRevLett.111.106403} {\bibfield  {journal} {\bibinfo
  {journal} {Phys. Rev. Lett.}\ }\textbf {\bibinfo {volume} {111}},\ \bibinfo
  {pages} {106403} (\bibinfo {year} {2013})}\BibitemShut {NoStop}%
\bibitem [{\citenamefont {Giannozzi}\ \emph {et~al.}(2009)\citenamefont
  {Giannozzi}, \citenamefont {Baroni}, \citenamefont {Bonini}, \citenamefont
  {Calandra}, \citenamefont {Car}, \citenamefont {Cavazzoni}, \citenamefont
  {Ceresoli}, \citenamefont {Chiarotti}, \citenamefont {Cococcioni},
  \citenamefont {Dabo}, \citenamefont {Corso}, \citenamefont {de~Gironcoli},
  \citenamefont {Fabris}, \citenamefont {Fratesi}, \citenamefont {Gebauer},
  \citenamefont {Gerstmann}, \citenamefont {Gougoussis}, \citenamefont
  {Kokalj}, \citenamefont {Lazzeri}, \citenamefont {Martin-Samos},
  \citenamefont {Marzari}, \citenamefont {Mauri}, \citenamefont {Mazzarello},
  \citenamefont {Paolini}, \citenamefont {Pasquarello}, \citenamefont
  {Paulatto}, \citenamefont {Sbraccia}, \citenamefont {Scandolo}, \citenamefont
  {Sclauzero}, \citenamefont {Seitsonen}, \citenamefont {Smogunov},
  \citenamefont {Umari},\ and\ \citenamefont {Wentzcovitch}}]{QEcode}%
  \BibitemOpen
  \bibfield  {author} {\bibinfo {author} {\bibfnamefont {P.}~\bibnamefont
  {Giannozzi}}, \bibinfo {author} {\bibfnamefont {S.}~\bibnamefont {Baroni}},
  \bibinfo {author} {\bibfnamefont {N.}~\bibnamefont {Bonini}}, \bibinfo
  {author} {\bibfnamefont {M.}~\bibnamefont {Calandra}}, \bibinfo {author}
  {\bibfnamefont {R.}~\bibnamefont {Car}}, \bibinfo {author} {\bibfnamefont
  {C.}~\bibnamefont {Cavazzoni}}, \bibinfo {author} {\bibfnamefont
  {D.}~\bibnamefont {Ceresoli}}, \bibinfo {author} {\bibfnamefont {G.~L.}\
  \bibnamefont {Chiarotti}}, \bibinfo {author} {\bibfnamefont {M.}~\bibnamefont
  {Cococcioni}}, \bibinfo {author} {\bibfnamefont {I.}~\bibnamefont {Dabo}},
  \bibinfo {author} {\bibfnamefont {A.~D.}\ \bibnamefont {Corso}}, \bibinfo
  {author} {\bibfnamefont {S.}~\bibnamefont {de~Gironcoli}}, \bibinfo {author}
  {\bibfnamefont {S.}~\bibnamefont {Fabris}}, \bibinfo {author} {\bibfnamefont
  {G.}~\bibnamefont {Fratesi}}, \bibinfo {author} {\bibfnamefont
  {R.}~\bibnamefont {Gebauer}}, \bibinfo {author} {\bibfnamefont
  {U.}~\bibnamefont {Gerstmann}}, \bibinfo {author} {\bibfnamefont
  {C.}~\bibnamefont {Gougoussis}}, \bibinfo {author} {\bibfnamefont
  {A.}~\bibnamefont {Kokalj}}, \bibinfo {author} {\bibfnamefont
  {M.}~\bibnamefont {Lazzeri}}, \bibinfo {author} {\bibfnamefont
  {L.}~\bibnamefont {Martin-Samos}}, \bibinfo {author} {\bibfnamefont
  {N.}~\bibnamefont {Marzari}}, \bibinfo {author} {\bibfnamefont
  {F.}~\bibnamefont {Mauri}}, \bibinfo {author} {\bibfnamefont
  {R.}~\bibnamefont {Mazzarello}}, \bibinfo {author} {\bibfnamefont
  {S.}~\bibnamefont {Paolini}}, \bibinfo {author} {\bibfnamefont
  {A.}~\bibnamefont {Pasquarello}}, \bibinfo {author} {\bibfnamefont
  {L.}~\bibnamefont {Paulatto}}, \bibinfo {author} {\bibfnamefont
  {C.}~\bibnamefont {Sbraccia}}, \bibinfo {author} {\bibfnamefont
  {S.}~\bibnamefont {Scandolo}}, \bibinfo {author} {\bibfnamefont
  {G.}~\bibnamefont {Sclauzero}}, \bibinfo {author} {\bibfnamefont {A.~P.}\
  \bibnamefont {Seitsonen}}, \bibinfo {author} {\bibfnamefont {A.}~\bibnamefont
  {Smogunov}}, \bibinfo {author} {\bibfnamefont {P.}~\bibnamefont {Umari}}, \
  and\ \bibinfo {author} {\bibfnamefont {R.~M.}\ \bibnamefont {Wentzcovitch}},\
  }\href {http://stacks.iop.org/0953-8984/21/i=39/a=395502} {\bibfield
  {journal} {\bibinfo  {journal} {Journal of Physics: Condensed Matter}\
  }\textbf {\bibinfo {volume} {21}},\ \bibinfo {pages} {395502} (\bibinfo
  {year} {2009})}\BibitemShut {NoStop}%
\bibitem [{\citenamefont {Giannozzi}\ \emph {et~al.}(2017)\citenamefont
  {Giannozzi}, \citenamefont {Andreussi}, \citenamefont {Brumme}, \citenamefont
  {Bunau}, \citenamefont {Nardelli}, \citenamefont {Calandra}, \citenamefont
  {Car}, \citenamefont {Cavazzoni}, \citenamefont {Ceresoli}, \citenamefont
  {Cococcioni}, \citenamefont {Colonna}, \citenamefont {Carnimeo},
  \citenamefont {Corso}, \citenamefont {de~Gironcoli}, \citenamefont {Delugas},
  \citenamefont {Jr}, \citenamefont {Ferretti}, \citenamefont {Floris},
  \citenamefont {Fratesi}, \citenamefont {Fugallo}, \citenamefont {Gebauer},
  \citenamefont {Gerstmann}, \citenamefont {Giustino}, \citenamefont {Gorni},
  \citenamefont {Jia}, \citenamefont {Kawamura}, \citenamefont {Ko},
  \citenamefont {Kokalj}, \citenamefont {Küçükbenli}, \citenamefont
  {Lazzeri}, \citenamefont {Marsili}, \citenamefont {Marzari}, \citenamefont
  {Mauri}, \citenamefont {Nguyen}, \citenamefont {Nguyen}, \citenamefont {de-la
  Roza}, \citenamefont {Paulatto}, \citenamefont {Poncé}, \citenamefont
  {Rocca}, \citenamefont {Sabatini}, \citenamefont {Santra}, \citenamefont
  {Schlipf}, \citenamefont {Seitsonen}, \citenamefont {Smogunov}, \citenamefont
  {Timrov}, \citenamefont {Thonhauser}, \citenamefont {Umari}, \citenamefont
  {Vast}, \citenamefont {Wu},\ and\ \citenamefont {Baroni}}]{QE-2017}%
  \BibitemOpen
  \bibfield  {author} {\bibinfo {author} {\bibfnamefont {P.}~\bibnamefont
  {Giannozzi}}, \bibinfo {author} {\bibfnamefont {O.}~\bibnamefont
  {Andreussi}}, \bibinfo {author} {\bibfnamefont {T.}~\bibnamefont {Brumme}},
  \bibinfo {author} {\bibfnamefont {O.}~\bibnamefont {Bunau}}, \bibinfo
  {author} {\bibfnamefont {M.~B.}\ \bibnamefont {Nardelli}}, \bibinfo {author}
  {\bibfnamefont {M.}~\bibnamefont {Calandra}}, \bibinfo {author}
  {\bibfnamefont {R.}~\bibnamefont {Car}}, \bibinfo {author} {\bibfnamefont
  {C.}~\bibnamefont {Cavazzoni}}, \bibinfo {author} {\bibfnamefont
  {D.}~\bibnamefont {Ceresoli}}, \bibinfo {author} {\bibfnamefont
  {M.}~\bibnamefont {Cococcioni}}, \bibinfo {author} {\bibfnamefont
  {N.}~\bibnamefont {Colonna}}, \bibinfo {author} {\bibfnamefont
  {I.}~\bibnamefont {Carnimeo}}, \bibinfo {author} {\bibfnamefont {A.~D.}\
  \bibnamefont {Corso}}, \bibinfo {author} {\bibfnamefont {S.}~\bibnamefont
  {de~Gironcoli}}, \bibinfo {author} {\bibfnamefont {P.}~\bibnamefont
  {Delugas}}, \bibinfo {author} {\bibfnamefont {R.~A.~D.}\ \bibnamefont {Jr}},
  \bibinfo {author} {\bibfnamefont {A.}~\bibnamefont {Ferretti}}, \bibinfo
  {author} {\bibfnamefont {A.}~\bibnamefont {Floris}}, \bibinfo {author}
  {\bibfnamefont {G.}~\bibnamefont {Fratesi}}, \bibinfo {author} {\bibfnamefont
  {G.}~\bibnamefont {Fugallo}}, \bibinfo {author} {\bibfnamefont
  {R.}~\bibnamefont {Gebauer}}, \bibinfo {author} {\bibfnamefont
  {U.}~\bibnamefont {Gerstmann}}, \bibinfo {author} {\bibfnamefont
  {F.}~\bibnamefont {Giustino}}, \bibinfo {author} {\bibfnamefont
  {T.}~\bibnamefont {Gorni}}, \bibinfo {author} {\bibfnamefont
  {J.}~\bibnamefont {Jia}}, \bibinfo {author} {\bibfnamefont {M.}~\bibnamefont
  {Kawamura}}, \bibinfo {author} {\bibfnamefont {H.-Y.}\ \bibnamefont {Ko}},
  \bibinfo {author} {\bibfnamefont {A.}~\bibnamefont {Kokalj}}, \bibinfo
  {author} {\bibfnamefont {E.}~\bibnamefont {Küçükbenli}}, \bibinfo {author}
  {\bibfnamefont {M.}~\bibnamefont {Lazzeri}}, \bibinfo {author} {\bibfnamefont
  {M.}~\bibnamefont {Marsili}}, \bibinfo {author} {\bibfnamefont
  {N.}~\bibnamefont {Marzari}}, \bibinfo {author} {\bibfnamefont
  {F.}~\bibnamefont {Mauri}}, \bibinfo {author} {\bibfnamefont {N.~L.}\
  \bibnamefont {Nguyen}}, \bibinfo {author} {\bibfnamefont {H.-V.}\
  \bibnamefont {Nguyen}}, \bibinfo {author} {\bibfnamefont {A.~O.}\
  \bibnamefont {de-la Roza}}, \bibinfo {author} {\bibfnamefont
  {L.}~\bibnamefont {Paulatto}}, \bibinfo {author} {\bibfnamefont
  {S.}~\bibnamefont {Poncé}}, \bibinfo {author} {\bibfnamefont
  {D.}~\bibnamefont {Rocca}}, \bibinfo {author} {\bibfnamefont
  {R.}~\bibnamefont {Sabatini}}, \bibinfo {author} {\bibfnamefont
  {B.}~\bibnamefont {Santra}}, \bibinfo {author} {\bibfnamefont
  {M.}~\bibnamefont {Schlipf}}, \bibinfo {author} {\bibfnamefont {A.~P.}\
  \bibnamefont {Seitsonen}}, \bibinfo {author} {\bibfnamefont {A.}~\bibnamefont
  {Smogunov}}, \bibinfo {author} {\bibfnamefont {I.}~\bibnamefont {Timrov}},
  \bibinfo {author} {\bibfnamefont {T.}~\bibnamefont {Thonhauser}}, \bibinfo
  {author} {\bibfnamefont {P.}~\bibnamefont {Umari}}, \bibinfo {author}
  {\bibfnamefont {N.}~\bibnamefont {Vast}}, \bibinfo {author} {\bibfnamefont
  {X.}~\bibnamefont {Wu}}, \ and\ \bibinfo {author} {\bibfnamefont
  {S.}~\bibnamefont {Baroni}},\ }\href
  {http://stacks.iop.org/0953-8984/29/i=46/a=465901} {\bibfield  {journal}
  {\bibinfo  {journal} {Journal of Physics: Condensed Matter}\ }\textbf
  {\bibinfo {volume} {29}},\ \bibinfo {pages} {465901} (\bibinfo {year}
  {2017})}\BibitemShut {NoStop}%
\bibitem [{\citenamefont {Dovesi}\ \emph
  {et~al.}(2018{\natexlab{a}})\citenamefont {Dovesi}, \citenamefont {Erba},
  \citenamefont {Orlando}, \citenamefont {Zicovich-Wilson}, \citenamefont
  {Civalleri}, \citenamefont {Maschio}, \citenamefont {Rérat}, \citenamefont
  {Casassa}, \citenamefont {Baima}, \citenamefont {Salustro},\ and\
  \citenamefont {Kirtman}}]{doi:10.1002/wcms.1360}%
  \BibitemOpen
  \bibfield  {author} {\bibinfo {author} {\bibfnamefont {R.}~\bibnamefont
  {Dovesi}}, \bibinfo {author} {\bibfnamefont {A.}~\bibnamefont {Erba}},
  \bibinfo {author} {\bibfnamefont {R.}~\bibnamefont {Orlando}}, \bibinfo
  {author} {\bibfnamefont {C.~M.}\ \bibnamefont {Zicovich-Wilson}}, \bibinfo
  {author} {\bibfnamefont {B.}~\bibnamefont {Civalleri}}, \bibinfo {author}
  {\bibfnamefont {L.}~\bibnamefont {Maschio}}, \bibinfo {author} {\bibfnamefont
  {M.}~\bibnamefont {Rérat}}, \bibinfo {author} {\bibfnamefont
  {S.}~\bibnamefont {Casassa}}, \bibinfo {author} {\bibfnamefont
  {J.}~\bibnamefont {Baima}}, \bibinfo {author} {\bibfnamefont
  {S.}~\bibnamefont {Salustro}}, \ and\ \bibinfo {author} {\bibfnamefont
  {B.}~\bibnamefont {Kirtman}},\ }\href {\doibase 10.1002/wcms.1360} {\bibfield
   {journal} {\bibinfo  {journal} {Wiley Interdisciplinary Reviews:
  Computational Molecular Science}\ }\textbf {\bibinfo {volume} {8}},\ \bibinfo
  {pages} {e1360} (\bibinfo {year} {2018}{\natexlab{a}})}\BibitemShut {NoStop}%
\bibitem [{\citenamefont {Dovesi}\ \emph
  {et~al.}(2018{\natexlab{b}})\citenamefont {Dovesi}, \citenamefont {Saunders},
  \citenamefont {Roetti}, \citenamefont {Orlando}, \citenamefont
  {Zicovich-Wilson}, \citenamefont {Pascale}, \citenamefont {Civalleri},
  \citenamefont {Doll}, \citenamefont {Harrison}, \citenamefont {Bush},
  \citenamefont {D’Arco}, \citenamefont {Llunel}, \citenamefont {Caus\`a},
  \citenamefont {No\"el}, \citenamefont {Maschio}, \citenamefont {Erba},
  \citenamefont {R\`erat},\ and\ \citenamefont {Casassa}}]{cryman}%
  \BibitemOpen
  \bibfield  {author} {\bibinfo {author} {\bibfnamefont {R.}~\bibnamefont
  {Dovesi}}, \bibinfo {author} {\bibfnamefont {V.~R.}\ \bibnamefont
  {Saunders}}, \bibinfo {author} {\bibfnamefont {C.}~\bibnamefont {Roetti}},
  \bibinfo {author} {\bibfnamefont {R.}~\bibnamefont {Orlando}}, \bibinfo
  {author} {\bibfnamefont {C.~M.}\ \bibnamefont {Zicovich-Wilson}}, \bibinfo
  {author} {\bibfnamefont {F.}~\bibnamefont {Pascale}}, \bibinfo {author}
  {\bibfnamefont {B.}~\bibnamefont {Civalleri}}, \bibinfo {author}
  {\bibfnamefont {K.}~\bibnamefont {Doll}}, \bibinfo {author} {\bibfnamefont
  {N.~M.}\ \bibnamefont {Harrison}}, \bibinfo {author} {\bibfnamefont {I.~J.}\
  \bibnamefont {Bush}}, \bibinfo {author} {\bibfnamefont {P.}~\bibnamefont
  {D’Arco}}, \bibinfo {author} {\bibfnamefont {M.}~\bibnamefont {Llunel}},
  \bibinfo {author} {\bibfnamefont {M.}~\bibnamefont {Caus\`a}}, \bibinfo
  {author} {\bibfnamefont {Y.}~\bibnamefont {No\"el}}, \bibinfo {author}
  {\bibfnamefont {L.}~\bibnamefont {Maschio}}, \bibinfo {author} {\bibfnamefont
  {A.}~\bibnamefont {Erba}}, \bibinfo {author} {\bibfnamefont {M.}~\bibnamefont
  {R\`erat}}, \ and\ \bibinfo {author} {\bibfnamefont {S.}~\bibnamefont
  {Casassa}},\ }\href {https://www.crystal.unito.it/Manuals/crystal17.pdf}
  {\bibfield  {journal} {\bibinfo  {journal} {CRYSTAL17 User's Manual}\ }
  (\bibinfo {year} {2018}{\natexlab{b}})}\BibitemShut {NoStop}%
\bibitem [{\citenamefont {Monkhorst}\ and\ \citenamefont
  {Pack}(1976)}]{PhysRevB.13.5188}%
  \BibitemOpen
  \bibfield  {author} {\bibinfo {author} {\bibfnamefont {H.~J.}\ \bibnamefont
  {Monkhorst}}\ and\ \bibinfo {author} {\bibfnamefont {J.~D.}\ \bibnamefont
  {Pack}},\ }\href {\doibase 10.1103/PhysRevB.13.5188} {\bibfield  {journal}
  {\bibinfo  {journal} {Phys. Rev. B}\ }\textbf {\bibinfo {volume} {13}},\
  \bibinfo {pages} {5188} (\bibinfo {year} {1976})}\BibitemShut {NoStop}%
\bibitem [{\citenamefont {Heyd}\ \emph {et~al.}(2003)\citenamefont {Heyd},
  \citenamefont {Scuseria},\ and\ \citenamefont
  {Ernzerhof}}]{doi:10.1063/1.1564060}%
  \BibitemOpen
  \bibfield  {author} {\bibinfo {author} {\bibfnamefont {J.}~\bibnamefont
  {Heyd}}, \bibinfo {author} {\bibfnamefont {G.~E.}\ \bibnamefont {Scuseria}},
  \ and\ \bibinfo {author} {\bibfnamefont {M.}~\bibnamefont {Ernzerhof}},\
  }\href {\doibase 10.1063/1.1564060} {\bibfield  {journal} {\bibinfo
  {journal} {The Journal of Chemical Physics}\ }\textbf {\bibinfo {volume}
  {118}},\ \bibinfo {pages} {8207} (\bibinfo {year} {2003})}\BibitemShut
  {NoStop}%
\bibitem [{\citenamefont {Heyd}\ \emph {et~al.}(2006)\citenamefont {Heyd},
  \citenamefont {Scuseria},\ and\ \citenamefont
  {Ernzerhof}}]{doi:10.1063/1.2204597}%
  \BibitemOpen
  \bibfield  {author} {\bibinfo {author} {\bibfnamefont {J.}~\bibnamefont
  {Heyd}}, \bibinfo {author} {\bibfnamefont {G.~E.}\ \bibnamefont {Scuseria}},
  \ and\ \bibinfo {author} {\bibfnamefont {M.}~\bibnamefont {Ernzerhof}},\
  }\href {\doibase 10.1063/1.2204597} {\bibfield  {journal} {\bibinfo
  {journal} {The Journal of Chemical Physics}\ }\textbf {\bibinfo {volume}
  {124}},\ \bibinfo {pages} {219906} (\bibinfo {year} {2006})}\BibitemShut
  {NoStop}%
\bibitem [{\citenamefont {Sophia}\ \emph {et~al.}(2013)\citenamefont {Sophia},
  \citenamefont {Baranek}, \citenamefont {Sarrazin}, \citenamefont
  {R{\'{e}}rat},\ and\ \citenamefont {Dovesi}}]{Sophia2013}%
  \BibitemOpen
  \bibfield  {author} {\bibinfo {author} {\bibfnamefont {G.}~\bibnamefont
  {Sophia}}, \bibinfo {author} {\bibfnamefont {P.}~\bibnamefont {Baranek}},
  \bibinfo {author} {\bibfnamefont {C.}~\bibnamefont {Sarrazin}}, \bibinfo
  {author} {\bibfnamefont {M.}~\bibnamefont {R{\'{e}}rat}}, \ and\ \bibinfo
  {author} {\bibfnamefont {R.}~\bibnamefont {Dovesi}},\ }\href {\doibase
  10.1080/01411594.2012.754442} {\bibfield  {journal} {\bibinfo  {journal}
  {Phase Transitions}\ }\textbf {\bibinfo {volume} {86}},\ \bibinfo {pages}
  {1069} (\bibinfo {year} {2013})}\BibitemShut {NoStop}%
\bibitem [{\citenamefont {Peralta}\ \emph {et~al.}(2006)\citenamefont
  {Peralta}, \citenamefont {Heyd}, \citenamefont {Scuseria},\ and\
  \citenamefont {Martin}}]{PhysRevB.74.073101}%
  \BibitemOpen
  \bibfield  {author} {\bibinfo {author} {\bibfnamefont {J.~E.}\ \bibnamefont
  {Peralta}}, \bibinfo {author} {\bibfnamefont {J.}~\bibnamefont {Heyd}},
  \bibinfo {author} {\bibfnamefont {G.~E.}\ \bibnamefont {Scuseria}}, \ and\
  \bibinfo {author} {\bibfnamefont {R.~L.}\ \bibnamefont {Martin}},\ }\href
  {\doibase 10.1103/PhysRevB.74.073101} {\bibfield  {journal} {\bibinfo
  {journal} {Phys. Rev. B}\ }\textbf {\bibinfo {volume} {74}},\ \bibinfo
  {pages} {073101} (\bibinfo {year} {2006})}\BibitemShut {NoStop}%
\bibitem [{\citenamefont {Heyd}\ \emph {et~al.}(2005)\citenamefont {Heyd},
  \citenamefont {Peralta}, \citenamefont {Scuseria},\ and\ \citenamefont
  {Martin}}]{doi:10.1063/1.2085170}%
  \BibitemOpen
  \bibfield  {author} {\bibinfo {author} {\bibfnamefont {J.}~\bibnamefont
  {Heyd}}, \bibinfo {author} {\bibfnamefont {J.~E.}\ \bibnamefont {Peralta}},
  \bibinfo {author} {\bibfnamefont {G.~E.}\ \bibnamefont {Scuseria}}, \ and\
  \bibinfo {author} {\bibfnamefont {R.~L.}\ \bibnamefont {Martin}},\ }\href
  {\doibase 10.1063/1.2085170} {\bibfield  {journal} {\bibinfo  {journal} {The
  Journal of Chemical Physics}\ }\textbf {\bibinfo {volume} {123}},\ \bibinfo
  {pages} {174101} (\bibinfo {year} {2005})}\BibitemShut {NoStop}%
\bibitem [{\citenamefont {Pernot}\ \emph {et~al.}(2015)\citenamefont {Pernot},
  \citenamefont {Civalleri}, \citenamefont {Presti},\ and\ \citenamefont
  {Savin}}]{Pernot2015}%
  \BibitemOpen
  \bibfield  {author} {\bibinfo {author} {\bibfnamefont {P.}~\bibnamefont
  {Pernot}}, \bibinfo {author} {\bibfnamefont {B.}~\bibnamefont {Civalleri}},
  \bibinfo {author} {\bibfnamefont {D.}~\bibnamefont {Presti}}, \ and\ \bibinfo
  {author} {\bibfnamefont {A.}~\bibnamefont {Savin}},\ }\href {\doibase
  10.1021/jp509980w} {\bibfield  {journal} {\bibinfo  {journal} {The Journal of
  Physical Chemistry A}\ }\textbf {\bibinfo {volume} {119}},\ \bibinfo {pages}
  {5288} (\bibinfo {year} {2015})}\BibitemShut {NoStop}%
\bibitem [{\citenamefont {Peintinger}\ \emph {et~al.}(2012)\citenamefont
  {Peintinger}, \citenamefont {Oliveira},\ and\ \citenamefont
  {Bredow}}]{Peintinger2012}%
  \BibitemOpen
  \bibfield  {author} {\bibinfo {author} {\bibfnamefont {M.~F.}\ \bibnamefont
  {Peintinger}}, \bibinfo {author} {\bibfnamefont {D.~V.}\ \bibnamefont
  {Oliveira}}, \ and\ \bibinfo {author} {\bibfnamefont {T.}~\bibnamefont
  {Bredow}},\ }\href {\doibase 10.1002/jcc.23153} {\bibfield  {journal}
  {\bibinfo  {journal} {Journal of Computational Chemistry}\ }\textbf {\bibinfo
  {volume} {34}},\ \bibinfo {pages} {451} (\bibinfo {year} {2012})}\BibitemShut
  {NoStop}%
\bibitem [{\citenamefont {Hummer}\ \emph {et~al.}(2009)\citenamefont {Hummer},
  \citenamefont {Harl},\ and\ \citenamefont {Kresse}}]{PhysRevB.80.115205}%
  \BibitemOpen
  \bibfield  {author} {\bibinfo {author} {\bibfnamefont {K.}~\bibnamefont
  {Hummer}}, \bibinfo {author} {\bibfnamefont {J.}~\bibnamefont {Harl}}, \ and\
  \bibinfo {author} {\bibfnamefont {G.}~\bibnamefont {Kresse}},\ }\href
  {\doibase 10.1103/PhysRevB.80.115205} {\bibfield  {journal} {\bibinfo
  {journal} {Phys. Rev. B}\ }\textbf {\bibinfo {volume} {80}},\ \bibinfo
  {pages} {115205} (\bibinfo {year} {2009})}\BibitemShut {NoStop}%
\bibitem [{\citenamefont {Odobescu}\ \emph {et~al.}(2017)\citenamefont
  {Odobescu}, \citenamefont {Maizlakh}, \citenamefont {Fedotov},\ and\
  \citenamefont {Zaitsev-Zotov}}]{PhysRevB.95.195151}%
  \BibitemOpen
  \bibfield  {author} {\bibinfo {author} {\bibfnamefont {A.~B.}\ \bibnamefont
  {Odobescu}}, \bibinfo {author} {\bibfnamefont {A.~A.}\ \bibnamefont
  {Maizlakh}}, \bibinfo {author} {\bibfnamefont {N.~I.}\ \bibnamefont
  {Fedotov}}, \ and\ \bibinfo {author} {\bibfnamefont {S.~V.}\ \bibnamefont
  {Zaitsev-Zotov}},\ }\href {\doibase 10.1103/PhysRevB.95.195151} {\bibfield
  {journal} {\bibinfo  {journal} {Phys. Rev. B}\ }\textbf {\bibinfo {volume}
  {95}},\ \bibinfo {pages} {195151} (\bibinfo {year} {2017})}\BibitemShut
  {NoStop}%
\end{thebibliography}%

\end{document}